\documentclass[12pt,aps,prd,showpacs,amsmath,amssymb]{revtex4}
\usepackage{slashed,amsmath,float}
\usepackage{mathrsfs}
\usepackage{amsfonts}
\usepackage{graphicx}
\usepackage{color}
\usepackage{ulem,bbm}
\usepackage{subfigure}
\usepackage{diagbox}
\usepackage{makecell}
\begin{document}
\title{Rare $\Lambda_b \rightarrow \Lambda l^+ l^- $  decay in the Bethe-Salpeter equation approach }

\author{Liang-Liang Liu $^{a}$}
\email{corresponding  author. liu06_04@sxnu.edu.cn}

\author{ Xian-Wei Kang $^{c}$}

\author{Zhen-Yang Wang $^{b}$}

\author{ Xin-Heng Guo $^{c}$}
\email{ corresponding  author. xhguo@bnu.edu.cn}

\affiliation{\footnotesize (a)~College of Physics and Information Engineering, Shanxi Normal University, Linfen 041004, People's Republic of China}
\affiliation {\footnotesize (b)~Physics Department, Ningbo University, Zhejiang, 315211, People's Republic of China}
\affiliation{\footnotesize (c)~College of Nuclear Science and Technology, Beijing Normal University, Beijing 100875, People's Republic of China}
\begin{abstract}
We study the rare decays $\Lambda_b \rightarrow \Lambda l^+ l^-~(l=e,\mu, \tau)$ in the Bethe-Salpeter equation approach.
We find that depending on the values of parameters in our model the branching ratio $Br(\Lambda_b \rightarrow \Lambda \mu^+ \mu^-)\times 10^{6}$ varies from $0.812$ to $1.445$ when $\kappa = 0.050 \sim 0.060$ GeV$^3$ and the binding energy $E_0=-0.14$ GeV while $Br(\Lambda_b \rightarrow \Lambda \mu^+ \mu^-)\times 10^{6}$ varies from $1.051$ to $1.098$ when $\kappa = 0.055$ Gev$^3$ and the binding energy $E_0$ changes from $-0.19$ to $-0.09$ GeV.
These results agree with the experimental data.
In the same parameter regions, we find that the branching ratio $Br(\Lambda_b \rightarrow \Lambda e^+ e^-(\tau^+ \tau^-) )\times 10^{6}$ varies in the range $0.660-1.028$ ($0.252-0.392$) and $0.749-1.098$ ($0.286-0.489$), respectively.

\end{abstract}

\maketitle

\section{Introduction}

In recent years, some interesting experimental results have been obtained in studies of rare decays of $b$ baryons induced by the $b \rightarrow s$ transition \cite{PRL107-201802, PLB725-25, PRL123-031801, JHEP09-146, JHEP06-115}.
The rare decay $\Lambda_b \rightarrow \Lambda \mu^+ \mu^-$ was observed by CDF \cite{PRL107-201802} and LHCb Collaboration \cite{PLB725-25}.
The first observation of the baryonic flavour changing neutral current decay $\Lambda_b \rightarrow \mu^- \mu^+$ by CDF Collaboration \cite{PRL107-201802} had a signal yield of $24\pm5$ events, corresponding to an absolute branching fraction $Br(\Lambda_b \rightarrow \Lambda \mu^+ \mu^-)=(1.73\pm 0.42$ (stat) $\pm$ (syst) $)\times 10^{-6}$ .
Following previous measurements, LHCb collaboration \cite{PLB725-25} gave a branching fraction of $Br(\Lambda_b \rightarrow \Lambda \mu^+ \mu^-)=(0.96\pm 0.16$(stat)$\pm 0.13$(syst)$\pm 0.21$(norm)$)\times 10^{-6}$ based on $78\pm12 ~\Lambda_b \rightarrow \mu^+ \mu^-$ events and updating the experimental data $d \Gamma(\Lambda_b \rightarrow \Lambda \mu^+ \mu^-)/dq^2 =(1.18^{+0.09}_{-0.08}\pm 0.036 \pm 0.27)\times 10^{-7}$GeV$^{-2}$ intergrating over $15<q^2<20$GeV$^2$ \cite{JHEP06-115}.
The first observation of the radiative decay $\Lambda_b \rightarrow \Lambda \gamma$ appeared in Ref. \cite{PRL123-031801} and the branching fraction was measured as $Br(\Lambda_b \rightarrow \gamma \Lambda)=(7.1\pm 1.56 \pm 0.6\pm 0.7)\times 10^{-6}$ based on $65\pm13 ~\Lambda_b \rightarrow \mu^+ \mu^-$ events with a significance of 5.6$\sigma$.
The analysis of the angular distribution of the decay $\Lambda_b \rightarrow \mu^+ \mu^-$ was done in \cite{JHEP09-146}, and first analysis of the differential fraction and the angular distribution of $\Lambda_b \rightarrow \mu^+ \mu^-$ were given in \cite{JHEP06-115}.
In the past several decades, there were many theoretical works to study the decay $\Lambda_b \rightarrow \Lambda \gamma$ \cite{JPG24-979, PTP102-645, EPJC59-847, PRD87-074031, PRD96-053006, PTEP073B04, PRD59-114022,PRD53-4946} and $\Lambda_b \rightarrow \Lambda \l^+ l^-$ \cite{PRD63-114024, PLB542-229, NPB649-168, PRD67-035007, EPJC38-283, NPB709-115, EPJC05-001, EPJC45-151, EPJC48-117, EPJC52-375, JHEP01-087, PRD81-056006, CTP58-872, PLB718-566, NPB863-398, PRD93-074501, EPJC78-230}.
Ref. \cite{JPG24-979} gave the branching fraction $Br(\Lambda_b \rightarrow \gamma \Lambda)=(1-4.5)\times 10^{-5}$ based on the experimental data \cite{PRL75-624}.
Ref. \cite{PTP102-645} gave the branching fraction $Br(\Lambda_b \rightarrow \gamma \Lambda)=0.23\times 10^{-5}$ in the Covariant Oscillator Quark Model.
Using QCD sum rules, Ref. \cite{PRD59-114022} gave $Br(\Lambda_b \rightarrow \gamma \Lambda)=(3.7\pm0.5) \times 10^{-5}$.
Following this work, considering the long distance effects, Ref. \cite{PRD63-114024} obtained the decay branching ratios $5.3\times 10^{-5}$ for $\Lambda_b\rightarrow \Lambda l^+ l^- (l=e,\mu)$ and $1.1\times 10^{-5}$ for $\Lambda_b\rightarrow \Lambda \tau^+ \tau^-$.
Using the decay form factors (FFs) from Ref. \cite{PRD59-114022}, there are many works to study the rare decay of $\Lambda_b \rightarrow \Lambda l^+ l^-$ \cite{PLB516-327, PLB542-229, NPB649-168, PRD67-035007, EPJC38-283, NPB709-115, EPJC48-117, EPJC52-375}.
In the relativistic quark model, Ref. \cite{PRD96-053006} obtained the branching fractions  $Br(\Lambda_b \rightarrow \Lambda l^+ l^-) \times 10^{-6}=1.07~(l=e), 1.05~(l=\mu), 0.26~(l=\tau)$.
However, in most of these works with the FFs of $\Lambda_b \rightarrow \Lambda$ being based on light-cone QCD sum rules and assumed to have the same shape, the results for the branching ratios of $\Lambda_b \rightarrow \Lambda l^+ l^-$ are different and do not agree with the experimental data.
One important way to search for new physics in b-physics is the analysis of rare $B$ decay model which are induced by the flavour changing neutral current (FCNC) transitions.
The FCNC transition is forbidden at the tree level in the standard model, and thus provides a good testing ground for new physics.
In order to use $\Lambda_b$ rare decays to search for new physics the $\Lambda_b \rightarrow \Lambda$ transition matrix must be determined more exact.

In the present work, we will use the Bethe-Salpeter (BS) equation to study this rare decay.
In our model, $\Lambda_{(b)}$ are described as a scalar diquark and quark bound systems, and then using the covariant instantaneous approximation the FFs of $\Lambda_b \rightarrow \Lambda$ will be calculated for giving the results for the $\Lambda_b \rightarrow \Lambda l^+ l^-$ decay branching ratios.
This paper is organized as follows. In Section II, we will establish the BS equation for $\Lambda_b$ and $\Lambda$.
In Section III we will derive the FFs for $ \Lambda_b \rightarrow \Lambda$ in the BS equation approach.
In Section IV the numerical results for the FFs and the decay branching ratios of $\Lambda_b \rightarrow \Lambda l^+ l^- $ will be given.
Finally, the summary and discussion will be given in Section V.

\section{BS EQUATION FOR $Q(ud)_{0 0}$ SYSTEM}\label{sec2}

In our work $\Lambda_b$ can be described as a $b(ud)_{00}$ system the first and second subscripts correspond to the spin and the isospin of $(ud)$, respectively) system.
The BS wave function of the $b(ud)_{0 0}$ system can be defined as the folowing \cite{CPC42-103106, PRD95-054001, PRD54-4629, PRD87-076013, PRD91-016006, PLB954-97, PRD86-056006}:
\begin{eqnarray}\label{chi-x}
  \chi(x_1,x_2,P) &=& \langle0|T\psi(x_1) \varphi(x_2)|P\rangle,
\end{eqnarray}
where $\psi(x_1)$ and $\varphi(x_2)$ are the field operators of the $b$-quark and $(ud)_{00}$ diquark, respectively, and $P$ is the momentum of $\Lambda_b$.
We use $M,~m,\text{and}~m_D$ to represent the masses of the $\Lambda_b$, the $b$-quark and the $(ud)$ diquark, respectively.
We define the BS wave function in momentum space:
\begin{eqnarray}\label{chi-f}
  \chi(x_1,x_2,P) = e^{i P X}\int \frac{d^4 p}{(2\pi)^4}e^{i p x} \chi_P(p),
\end{eqnarray}
where $X= \lambda_1 x_1+\lambda_2 x_2$ is the coordinate of mass center,
$\lambda_1=\frac{m }{m +m_D} $, $\lambda_2=\frac{m_D}{m +m_D} $, and $x= x_1-x_2$.
In momentum space, the BS equation for the $b(ud)_{00}$ system satisfies the homogeneous integral equation \cite{CPC42-103106, PRD95-054001, PRD54-4629, PRD87-076013, PRD91-016006, PLB954-97, PRD86-056006}
\begin{eqnarray}\label{chi-p}
\chi_P(p) = i S_F(p_1)\int \frac{d^4 p}{(2 \pi)^4}  [ I\otimes I V_1(p,q)+ \gamma_\mu \otimes \Gamma^\mu V_2(p,q)   ]\chi_P(q)S_D(p_2),
\end{eqnarray}

\begin{figure}[!ht]
\begin{center}
\includegraphics[width=7.5cm] {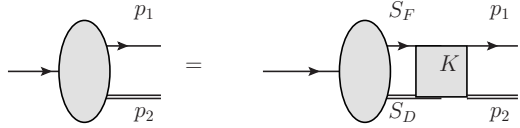}
\caption{The BS equation for $b(ud)_{00}$ system in momentum space (K is the interaction kernel).}\label{BSE}
\end{center}
\end{figure}
where the quark momentum $p_1=\lambda_1 P+p$ and the diquark momentum $p_2=\lambda_2 P-p$, $S_F(p_1)$ and $S_D(p_2)$ are propagators of the quark and the scalar diquark, respectively,  $\Gamma^\mu=(p_2+q_2)^\mu \frac{\alpha_{seff}Q_0^2}{Q^2+Q^2_0}$ is introduced to describe the structure of the scalar diquark \cite{PRD54-4629, JPG24-979, PRD22-2157}.
By analyzing the electromagnetic FFs of proton, it was found that $Q_0^2=3.2$ GeV$^2$ can lead to consistent results with the experimental data \cite{JPG24-979}.
$V_1$ and  $V_2$ are the scalar confinement and one-gluon-exchange terms, respectively.
Generally, the $b(ud)_{00}$ system needs two scalar functions to describe the BS wave function \cite{CPC42-103106, PRD95-054001, PRD91-016006}
\begin{eqnarray}
  \chi_P(p) &=& (f_1(p_t^2)+\slashed{p}_t f_2(p_t^2))u(P),
\end{eqnarray}
where $f_i, (i=1,2)$ are the Lorentz-scalar functions of $p_t^2$, $u(P)$ is the spinor of $\Lambda_b$, $p_t$ is the transverse projection of the relative momenta along the momentum $P$, $p_t^\mu = p^\mu-(v\cdot p) v^\mu$ and $p_l= \lambda_2 M- v \cdot p  $ (where we have defined $v^\mu=P^\mu/M$).
Motivated by the potential model, $V_1$ and $V_2$ have the following forms in the covariant instantaneous approximation ($p_l=q_l$) \cite{PRD54-4629, PRD87-076013, PRD86-056006, PRD76-056004}:
\begin{eqnarray}\label{V1}
  \tilde{V}_1(p_t-q_t) = \frac{8 \pi \kappa}{[(p_t-q_t)^2+\mu^2]^2} - (2\pi)^2\delta^3(p_t-q_t)\int \frac{d^3k}{(2\pi)^3} \frac{8 \pi \kappa}{(k^2+\mu^2)^2},
\end{eqnarray}

\begin{eqnarray}\label{V2}
  \tilde{V_2} (p_t-q_t)&=&- \frac{16 \pi }{3}\frac{\alpha_{seff} }{(p_t-q_t)^2+\mu^2},
\end{eqnarray}
where $q_t$ is the transverse projection of the relative momenta along the momentum $P$ and defined as $ q_t^\mu = q^\mu -(v \cdot q) v^\mu$, $q_l=\lambda_2 M -v \cdot q$. The second term of $\tilde{V}_1$ is introduced to avoid infrared divergence at the point $ p_t=q_t$, $\mu$ is a small parameter to avoid infrared divergence.
The parameters $\kappa$ and $\alpha_{seff}$ are related to scalar confinement and the one-gluon-exchange diagram, respectively.

The quark and diquark propagators can be written as the following:
\begin{eqnarray}\label{SF}
  S_F(p_1) = i \slashed{v} \bigg[ \frac{\Lambda_q^+ }{ M -p_l -\omega_q  +i \epsilon} +\frac{\Lambda_q ^-}{ M -p_l +\omega  -i \epsilon}\bigg],
\end{eqnarray}
\begin{eqnarray}\label{SD}
  S_D(p_2)  = \frac{i}{2 \omega_D} \bigg[\frac{1}{ p_l-\omega_D+i \epsilon} -\frac{1}{ p_l+ \omega_D-i\epsilon}\bigg],
\end{eqnarray}
where $\omega_q = \sqrt{m^2-p_t^2}~\text{and}~\omega_D = \sqrt{m_D^2-p_t^2} $.
$\Lambda^\pm_q=1/2 \pm  \slashed{v}(\slashed{p}_t+m)/(2\omega_q)$ are the projection operators which satisfy the relations, $\Lambda_q^\pm \Lambda_q^\pm = \Lambda^\pm_q,~ \Lambda^\pm_q \Lambda^\mp_q = 0$.
At the order of $\frac{1}{m}$\cite{PRD54-4629}, the quark propagator can be written as
\begin{eqnarray}\label{SF-HQ}
  S_F(p_1) = i \frac{ 1+ \slashed{v}  }{  2 (E_0+m_D -p_l+ i \epsilon) },
\end{eqnarray}
where $E_0=M-m-m_D$ is the binding energy. In general,  $E_0$ is about $-0.14 \pm 0.05$ GeV \cite{PRD91-016006}.
 Then we can get $\kappa$ is about $0.05\pm0.01$ GeV$^3$ for $\Lambda_b$ \cite{PLB954-97}.
Defining $\tilde{f}_{1(2)}=\int \frac{d p_l}{2 \pi}f_{1(2)}$, and using the covariant instantaneous approximation, $p_l=q_l$,  the scalar BS wave functions satisfy the coupled integral equation

\begin{eqnarray}\label{eig1}
&& \tilde{f}_1(p_t) =\int \frac{d^3q_t}{(2\pi)^3} M_{11}(p_t,q_t) \tilde{f}_1(q_t)+  M_{12}(p_t,q_t) \tilde{f}_2(q_t) ,
\end{eqnarray}

\begin{eqnarray}\label{eig2}
 &&\tilde{f}_2(p_t) = \int \frac{d^3q_t}{(2\pi)^3}  M_{21}(p_t,q_t) \tilde{f}_1(q_t) +  M_{22}(p_t,q_t) \tilde{f}_2(q_t),
\end{eqnarray}
where
\begin{eqnarray}
M_{11}(p_t,q_t)=\frac{(\omega_q  +m ) (\tilde{V}_1+ 2 \omega_D \tilde{V}_2)-   p _t \cdot ( p _t+ q _t) \tilde{V}_2}{4 \omega_D \omega_q(-M + \omega_D+ \omega_q)} - \nonumber\\ \frac{(\omega_q -m )(\tilde{V}_1- 2\omega_D \tilde{V}_2)+   p _t\cdot( p _t+ q _t)  \tilde{V}_2}{4 \omega_D \omega_c(M + \omega_D+ \omega_q)},
\end{eqnarray}

\begin{eqnarray}
M_{12}(p_t,q_t)=\frac{-  (\omega_q+m ) ( q _t + p _t)\cdot q_t\tilde{V}_2 +  p _t\cdot q_t(\tilde{V}_1- 2 \omega_D \tilde{V}_2)}{4 \omega_D \omega_c(-M + \omega_D+ \omega_c)}- \nonumber\\ \frac{(m - \omega_q )  ( q _t + p _t)\cdot q _t \tilde{V}_2 -   p _t\cdot q _t  (\tilde{V}_1+ 2\omega_D \tilde{V}_2)}{4 \omega_D \omega_q(M + \omega_D+ \omega_q)},
\end{eqnarray}

\begin{eqnarray}
M_{21}(p_t,q_t)= \frac{(\tilde{V}_1+ 2 \omega_D \tilde{V}_2)-( -\omega_q+m ) \frac{( p _t+ q _t) \cdot p _t }{ p^2_t }\tilde{V}_2}{4 \omega_D \omega_q(-M + \omega_D+ \omega_q)}    - \nonumber\\
\frac{- (\tilde{V}_1- 2\omega_D \tilde{V}_2)+(\omega_q  + m )\frac{( p _t+ q _t)\cdot p _t }{ p^2_t } \tilde{V}_2)}{4 \omega_D \omega_q(M + \omega_D+ \omega_q)},
\end{eqnarray}

\begin{eqnarray}
M_{22}(p_t,q_t)=  \frac{(m  -\omega_q)( \tilde{V}_1+ 2  \omega_D \tilde{V}_2）  ) \frac{ p_t \cdot q_t}{ p^2_t } - (  q^2_t+  p_t \cdot q_t) \tilde{V}_2}{4 \omega_D \omega_q(-M + \omega_D+ \omega_q)} - \nonumber\\
\frac{ (m +\omega_q) (-\tilde{V}_1- 2 \omega_D \tilde{V}_2）) \frac{p_t \cdot q_t}{p^2_t} + (  q^2_t+  p_t \cdot q_t)\tilde{V}_2)}{4 \omega_D \omega_q(M + \omega_D+ \omega_q)}.
\end{eqnarray}

For $\Lambda_b$, when $\frac{1}{m_b}\rightarrow 0$ and considering Dirac equation for $\Lambda_b$ we have

\begin{eqnarray}\label{eig3}
  \phi(p) &=& -\frac{i}{(E_0+m_D-p_l+i \epsilon)( p_l ^2-\omega^2_D)}\int \frac{d^4 q }{(2\pi)^4}(\tilde{V}_1+2  p_l \tilde{V}_2)\phi(q).
\end{eqnarray}

The BS wave function of $\Lambda_b$ was given in the previous work \cite{PRD54-4629} and has the form $\chi_P (v)=\phi (p)u_{\Lambda_b}(v,s)$, where $\phi(p)$ is the scalar BS wave function.

Generally, the BS wave function can be normalized under the condition of the covariant instantaneous approximation \cite{PRD76-056004}:
\begin{eqnarray}\label{BSNOR}
  i \delta^{i_1 i_2}_{j_1 j_2} \int \frac{d^4 q d^4 p}{(2\pi)^8}\bar{\chi}_P(p,s)\left[\frac{\partial}{\partial P_0}I_p(p,q)^{i_1 i_2 j_2 j_1}\right]\chi_P(q,s^\prime) =\delta_{s s^\prime},
\end{eqnarray}
where $i_{1(2)}$ and $j_{1(2)}$ represent the color indices of the quark and the diquark, respectively, $s^{(\prime)}$ is the spin index of the baryon $\Lambda_b$, $I_p(p,q)^{i_1 i_2 j_2 j_1}$ is the inverse of the four-point propagator written as follows
\begin{eqnarray}\label{IPNOR}
  I_p(p,q)^{i_1 i_2 j_2 j_1} =\delta^{i_1 j_1}\delta^{i_2 j_2} (2 \pi)^4 \delta^4(p-q)S^{ -1 }_F(p_1)S^{ -1 }_D(p_2).\nonumber\\
\end{eqnarray}

\section{Matrix element of $\Lambda_b \rightarrow \Lambda l^+ l^-$ decay}

In this section, we derive the matrix element of $\Lambda_b\rightarrow\Lambda l^+ l^-$ in the BS equation approach.
At the quark level, $\Lambda_b\rightarrow\Lambda l^+l^-$ is described by the $b\rightarrow s l^+l^-$ transition.
The effective Hamiltonian describing the electroweak penguin and weak box diagrams related to this transition is given by

\begin{eqnarray}
\mathcal{H} = \frac{G_F\alpha}{\sqrt{2}\pi}V_{tb}V^*_{ts}\bigg\{ \bar{s}\bigg[C^{eff}_9 \gamma_{\mu}P_L -i C^{eff}_{7}\frac{2 m_b\sigma_{\mu\nu} q^{\mu}}{q^2}P_R \bigg]b(\bar{l}\gamma_{\mu}l)+C_{10}(\bar{s}\gamma_{\mu}P_L b) (\bar{l}\gamma^{\mu}\gamma_5l) \bigg\},
\end{eqnarray}
where $G_F$ and $\alpha$ are to the Fermi coupling constant and the electromagnetic coupling constant, respectively, $ P_{R,L}= (1\pm \gamma_5)/2,$  $q$ is the total momentum of the lepton pair and $C_i~(i=7,~9,~10,)$ are the Wilson coefficients.
The amplitude of the decay $\Lambda_b \rightarrow \Lambda l^+ l^-$ is obtained by calculating the matrix element of effective Hamiltonian for the $b \rightarrow s l^+ l^-$ transition between the initial and final states, $\langle \Lambda | \mathcal{H}|\Lambda_b\rangle$. The matrix element can be parameterized in terms of the FFs as the following:

\begin{eqnarray} \label{FFs}
 \langle\Lambda(P',s')\arrowvert \bar{s}\gamma_{\mu}b\arrowvert\Lambda_b(P,s)\rangle = \bar{u}_{\Lambda}(P',s')(g_1\gamma^\mu+ ig_2\sigma_{\mu\nu}p^{\nu}+g_3p_\mu)u_{\Lambda_b}(P,s),\nonumber\\
 \langle\Lambda(P',s')\arrowvert \bar{s}\gamma_{\mu}\gamma_{5}b\arrowvert\Lambda_b(P,s)\rangle= \bar{u}_{\Lambda}(P',s')(t_1\gamma^\mu+it_2\sigma_{\mu\nu}p^{\nu}+t_3p^\mu)\gamma_5u_{\Lambda_b}(P,s),\nonumber\\
 \langle\Lambda(P',s')\arrowvert \bar{s}i\sigma^{\mu\nu}q^{\nu}b\arrowvert\Lambda_b(P,s)\rangle= \bar{u}_{\Lambda}(P',s')(s_1\gamma^\mu+is_2\sigma_{\mu\nu}q^{\nu}+s_3q^\mu)u_{\Lambda_b}(P,s),\nonumber\\
 \langle\Lambda(P',s')\arrowvert \bar{s}i\sigma^{\mu\nu}\gamma_5q^{\nu}b\arrowvert\Lambda_b(P,s)\rangle= \bar{u}_{\Lambda}(P',s')(d_1\gamma^\mu+id_2\sigma_{\mu\nu}q^{\nu}+d_3q^\mu)\gamma_5u_{\Lambda_b}(P,s),
\end{eqnarray}

\noindent where $q=P-P'$ is the momentum transfer, and $g_i$, $t_i$, $s_i $, $d_i $ ($i=1,2$ and 3) are various form factors which are Lorentz scalar functions of $q^2$.
Considering the spin symmetry on the $b$ quark in the limit
$m_b\rightarrow \infty $, the matrix elements in Eq. (\ref{FFs}) can be rewritten as

\begin{eqnarray}\label{FFs-HQET}
  \langle\Lambda(P',s')\arrowvert \bar{s}\Gamma_{\mu} b\arrowvert \Lambda_b(v,s)\rangle =\bar{u}_{\Lambda}(P',s')(F_{1}(\omega)+F_2(\omega)\slashed{v})\Gamma^{\mu}u_{\Lambda_b}(v,s),
\end{eqnarray}

\noindent where $\Gamma_{\mu}$ represent $\gamma_{\mu}$, $\gamma_{\mu}\gamma_5$, $i\sigma_{\mu\nu}q^{\nu}$, and $i\sigma_{\mu\nu}\gamma_5q^{\nu}$. $F_i$ ($i=1, 2$) can be expressed as functions solely of $\omega=v\cdot P'/m_{\Lambda}$, which is the energy of the $\Lambda$ baryon in the $\Lambda_b$ rest frame.
In the pole formulae for the extrapolation to $q^2=0$ in the decay $\Lambda_b \rightarrow \Lambda \gamma$ we have $F_1(0)=0.45$ (monopole) and $F_1(0)=0.22$ (dipole) \cite{JPG24-979}, while the author of Ref. \cite{JPG24-979} combined the CLEO data from Ref. \cite{PRL75-624} to get $F_1(q^2_{max})=1.21$ ignoring the mass of $\Lambda$ baryon.
Lattice QCD (LQCD) gives $F_1(q^2_{max} )\approx 1.25$ at the leading order in the heavy quark effective theory \cite{PRD87-074502}.
In Ref. \cite{PRD60-014003} it was assumed $F_2=0$.
The QCD sum rules analysis obtained that $F_1=0.50\pm0.03$ and $F_2=-0.1\pm0.03$ at the point $E_0=(m_{\Lambda_b}^2+m_\Lambda^2)/(2  m_{ \Lambda_b}) =2.93$ GeV.
Therefore, we expect $F_1(q^2_{max}) < 1.5$, considering the correction of $\Lambda_{QCD}/m_b$.
The ratio $R= F_2/F_1 =-0.35\pm0.04$ (stat) $\pm0.04$ (syst) has been previously  measured by the CLEO Collaboration using experimental data for the semileptonic decay $\Lambda_c \rightarrow \Lambda e^+ \nu_e$ with the invariant mass in the range from $m_\Lambda$ to $m_{\Lambda_c}$, assuming the same shape for $F_1$ and $F_2$ and ignoring the $\Lambda_{QCD}/m_c$ corrections \cite{PRL94-191801}.
In Ref. \cite{PRD59-114022} $R=-0.42~(-0.83)$ was given at $q^2=q^2_{max}(q^2=0)$, and in Ref. \cite{PLB516-327} $R(0) \equiv -0.17$ and $R(q_{max}^2=m^2_{\Lambda_c})=-0.44$ were obtained.
However, according to pQCD scaling law, the FFs should have different shapes for large $q^2$ \cite{PRD11-1309,PRD22-2157,PPNP59-694}, therefore, expect $R(q^2)\propto -1/q^2$, which agrees with Ref. \cite{PRD53-4946}.
Using the experimental data \cite{PRL94-191801}, we have estimated the value of $R(q^2_{max}=m_{\Lambda_b}-m_\Lambda)^2)$ and found it should be from $-1.12$ to $-0.7$ approximately.
Considering Ref. \cite{PRD59-114022}, we let $R(q^2_{max})$ to vary from $-0.83$ to $-0.7$.

Comparing Eq. (\ref{FFs}) with Eq. (\ref{FFs-HQET}), we obtain the following relations:
\begin{eqnarray}
 & & g_1~=~t_1~=~s_2~=~d_2~=~\bigg(F_1+\sqrt{r}F_2\bigg),\nonumber\\
 & & g_2~=~t_2~=g_3~=~t_3~=~\frac{1}{m_{\Lambda_{b}}}F_2, \nonumber\\
 & & s_3~=~  F_2 (\sqrt{r}-1),~ d_3~=~ F_2(\sqrt{r}+1), \nonumber\\
 & & s_1 ~=~ d_1~=~ F_2 m_{\Lambda_b}  (1+r-2\sqrt{r}\omega),
\end{eqnarray}
where $r=m_{\Lambda}^2/m_{\Lambda_b}^2$.
The transition matrix for $\Lambda_b\rightarrow \Lambda$ can be expressed in terms of the BS wave function of $\Lambda_b$ and $\Lambda$,
\begin{eqnarray}\label{FFs-BS}
  \langle\Lambda(P',s')|\bar{s}\Gamma_{\mu}b|\Lambda_b(P,s)\rangle =\int\frac{d^4p}{(2\pi)^4} \bar{\chi}_{P'}^{\Lambda}(p')\Gamma_{\mu}\chi_P^{\Lambda_b}(p)S^{-1}_D(p_2).
\end{eqnarray}

Define
\begin{eqnarray}
  \int \frac{d^4p}{(2 \pi)^4} f_1(p^\prime) \phi(p) S^{-1}_D(p_2)=k_1(\omega), \nonumber\\
  \int \frac{d^4p}{(2 \pi)^4} f_2(p^\prime)p_{t\mu}^\prime \phi(p) S^{-1}_D(p_2)=k_2(\omega) v_{\mu} + k_3(\omega) v^\prime_{\mu},
\end{eqnarray}
where  $v^\prime = P^\prime/m_\Lambda$, then we find the following relations when $\omega \neq 1$
\begin{eqnarray}
  k_3 &=& - \omega k_2, \nonumber\\
  k_2 &=& \frac{1}{1-\omega^2} \int \frac{d^4 p}{(2\pi)^4} f_2(p^\prime) p^\prime_t \cdot v \phi(p) S^{-1}_D,
\end{eqnarray}
and
\begin{eqnarray}
  F_1 &=& k_1- \omega k_2 , \nonumber\\
  F_2&=&k_2.
\end{eqnarray}

The differential decay rate is obtained as the flowing:
\begin{eqnarray}
 \mathcal{M}(\Lambda_b\rightarrow \Lambda l^{+} l^{-})&=&\frac{G_F}{ \sqrt{2}\pi}\times \lambda_t\big[\bar{l}\gamma_{\mu}l\{\bar{u}_{\Lambda}[\gamma_{\mu}(A_1P_R +B_1P_L)+i\sigma^{\mu\nu}p_{\nu}(A_2 P_R +B_2P_L)]u_{\Lambda_b}\} \nonumber\\
&+&\bar{l}\gamma_{\mu}\gamma_5l\{\bar{u}_{\Lambda}[\gamma^{\mu}(D_1P_R +E_1P_L)+i\sigma^{\mu\nu}p_{\nu}(D_2P_R+E_2P_L)\nonumber\\ &+&p^{\mu}(D_3P_R+E_3P_L)]u_{\Lambda_b}\}\big],
\end{eqnarray}
where the parameters $A_i$, $B_i$ and $D_j$, $E_j$ ($i=1,2$ and $j=1,2,3$) are defined as
\begin{eqnarray}
&&A_i=\frac{1}{2}\bigg\{C^{eff}_{9}(g_i-t_i)-\frac{2C^{eff}_7 m_b}{p^2}(d_i +s_i )\bigg\},\nonumber\\
& &B_i = \frac{1}{2}\bigg\{C^{eff}_{9}(g_i+t_i) - \frac{2C^{eff}_7m_b}{p^2}(d_i -s_i )\bigg\}, \nonumber\\
& &D_j = \frac{1}{2}C_{10}(g_j-t_j), ~E_j=\frac{1}{2}C_{10}(g_j+t_j).
\end{eqnarray}

In the physical region$(4m^2_l\leq q^2\leq (m_{\Lambda_b}-m_{\Lambda})^2)$, the decay rate of $\Lambda_b\rightarrow\Lambda l^+l^-$ is obtained as

\begin{eqnarray}
\frac{d\Gamma}{dq^2}=\frac{G^2_F\alpha^2}{2^{13}\pi^5m_{\Lambda_b}} |V_{tb}V^*_{ts}|^2v_l\sqrt{\lambda(1,r,s)} \mathcal{M}(s)  ,
\end{eqnarray}
where  $s=q^2/m^2_{\Lambda_b}(q^2 = m^2_{\Lambda_b}+m^2_\Lambda - 2 m_{\Lambda_b} m_\Lambda \omega)$, $ \lambda(1,r,s)=1+r^2+s^2-2r-2s-2rs,$ and  $v_l=\sqrt{1-\frac{4m^2_l}{q^2}}$  is the lepton velocity.
The decay amplitude is given as \cite{EPJC45-151}

\begin{eqnarray}
   \mathcal{M}(s) &=& \mathcal{M}_0(s) +\mathcal{M}_2(s),
\end{eqnarray}
where
\begin{eqnarray}
  \mathcal{M}_0(s)&&=32m^2_l m^4_{\Lambda_b}s(1+r-s)(|D_3|^2+|E_3|^2) \nonumber\\
  &&64m^2_lm^3_{\Lambda_b}(1-r-s)Re(D^*_1E_3+D_3E^*_1)\nonumber\\
& &+64m^2_{\Lambda_b}\sqrt{r}(6m^2_l-M^2_{\Lambda_b}s)Re(D_1^*E_1)\nonumber\\
&& 64m^2_lm^3_{\Lambda}\sqrt{r}\big(2m_{\Lambda_b}s Re(D^*_3E_3) +(1-r+s)Re(D^*_1D_3+E^*_1E_3)\big)\nonumber\\
&&+32m^2_{\Lambda}(2m^2_l+m^2_{\Lambda}s)\bigg\{(1-r+s)m_{\Lambda_b}\sqrt{r}Re(A^*_1A_2+B^*_1B_2)\nonumber\\
& &-m_{\Lambda_b}(1-r-s)Re(A^*_1B_2+A^*_2B_1)  -2\sqrt{r}\big(Re(A^*_1B_1)+m^2_{\Lambda}s Re(A^*_2B_2)\big) \bigg \}\nonumber\\
& &+ 8 m^2_{\Lambda_b}\bigg[4m^2_l(1+r-s)+m^2_{\Lambda_b}((1+r)^2- s^2)\bigg](|A_1|^2+|B_1|^2)\nonumber\\
&&+8m^4_{\Lambda_b}\bigg\{4m^2_l[\lambda+(1+r-s)s]+m^2_{\Lambda_b}s[(1-r)^2-s^2]\bigg\}(|A_2|^2+|B_2|^2) \nonumber\\
& & - 8m^2_{\Lambda_b}\bigg\{4m^2_l(1+r-s)-m_{\Lambda_b}[(1-r)^2-s^2]\bigg\} (|D_1|^2+|E_1|^2) \nonumber\\
&&+ 8m^5_{\Lambda_b}sv^2\bigg\{-8m_{\Lambda_b}s\sqrt{r}Re(D^*_2E_2) +4(1-r+s)\sqrt{r}Re(D^*_1D_2+E^*_1E_2)\nonumber\\
&& -4(1-r-s) Re(D^*_1E_2+D^*_2E_1)+m_{\Lambda_b}[(1-r)^2-s^2] (|D_2|^2+|E_2|^2)\bigg\},
\end{eqnarray}
\begin{eqnarray}
  \mathcal{M}(s) &=& 8m^6_{\Lambda_b}s v_l^2\lambda(|A_2|^2+|B_2|^2+|C_2|^2+|D_2|^2) \nonumber\\ &- &8 m^4_{\Lambda_b}v_l^2\lambda(|A_1|^2+|B_1|^2+|C_1|^2+|D_1|^2).
\end{eqnarray}

\section{Numerical analysis}

In order to analyze the decay rate and branching ratio, we use the following the numerical values: for the Wilson coefficients,  $C^{eff}_7=-0.313$, $C^{eff}_9=4.334$, $C_{10}=-4.669$ \cite{JHEP10-118, PRD79-074007, EPJC40-565}, for the masses of baryons, $m_{\Lambda_b}=5.62$ GeV, $m_\Lambda=1.116$ GeV \cite{PRD98-030001}, while for the masses of quark, $m_b=5.02$ GeV and $m_s=0.516$ GeV \cite{PRD95-054001, PRD87-076013, PRD91-016006}.
The variable $\omega$ varies from $1$ to $2.617,~2.614$, and $1.617$ for $e,~\mu$, and $\tau$, respectively.

Solving Eqs. (\ref{eig1}) and (\ref{eig2}) for $\Lambda$ with the parameters we have taken,  one can get the numerical solutions of BS wave functions.
For $\Lambda_b$ we need to solve Eq. (\ref{eig3}).
In Table. \ref{TB1}, we give the values of $\alpha_s$ with different binding energy $E_0$  and different $\kappa $ for $\Lambda$.
In Table. \ref{TB2}, we give the values of $\alpha_s$ with different binding energy $E_0$  and different $\kappa $ for $\Lambda_b$.
It can be seen from Tables. \ref{TB1} and \ref{TB2} that the dependence of $\alpha_{seff}$ on the parameters $\kappa$ and $E_0$ for $\Lambda$ is obviously stronger than that for $\Lambda_b$.

\begin{table}[!htb]
\centering  
\begin{tabular}{c||c|c|c|c|c|c|c|c|c|c|c}  
\hline
 \diagbox{$ E_0$ }{$\alpha_{seff}$}{$\kappa \times 10^{3}$ }& 40  & 42 & 44 & 46  & 48 & 50 & 52 & 54  & 56  & 58 & 60 \\ \hline \hline
 -0.19 &0.616 &0.611 &0.661 &0.606& 0.601&0.596&0.592&0.588&0.584&0.580&0.577 \\         
 -0.14  &0.576 &0.570 &0.566 &0.561&0.557&0.553&0.549&0.546&0.542&0.539&0.536 \\        
-0.09 &0.521 &0.517 &0.513 &0.509&0.506&0.503&0.500&0.497&0.495&0.492&0.490\\ \hline \hline
\end{tabular}
\caption{The values of $\alpha_{seff}$  for $\Lambda$ (the units of $E_0$ and $\kappa$ are GeV and GeV$^3$, respectively).}\label{TB1}
\end{table}

\begin{table}[!htb]
\centering  
\begin{tabular}{c||c|c|c|c|c|c|c|c|c|c|c}  
\hline
 \diagbox{$ E_0$ }{$\alpha_s$}{$\kappa \times 10^{3}$ }& 40  & 42 & 44 & 46  & 48 & 50 & 52 & 54  & 56  & 58 & 60 \\ \hline \hline
 -0.19 &0.806 &0.808 &0.809 &0.796& 0.811&0.812&0.814&0.815&0.817&0.818&0.819 \\         
 -0.14  &0.770 &0.772 &0.774 &0.776&0.777&0.779&0.781&0.783&0.785&0.786&0.788 \\        
-0.09 &0.729 &0.732 &0.735 &0.737&0.713&0.740&0.742&0.744&0.747&0.749&0.751\\ \hline \hline
\end{tabular}
\caption{The values of $\alpha_{seff}$  for $\Lambda_b$ (the units of $E_0$ and $\kappa$ are GeV and GeV$^3$, respectively).}\label{TB2}
\end{table}

\begin{figure}[!htb]
\begin{center}
\begin{minipage}[t]{0.45\linewidth}
 \includegraphics[width=7.0cm]{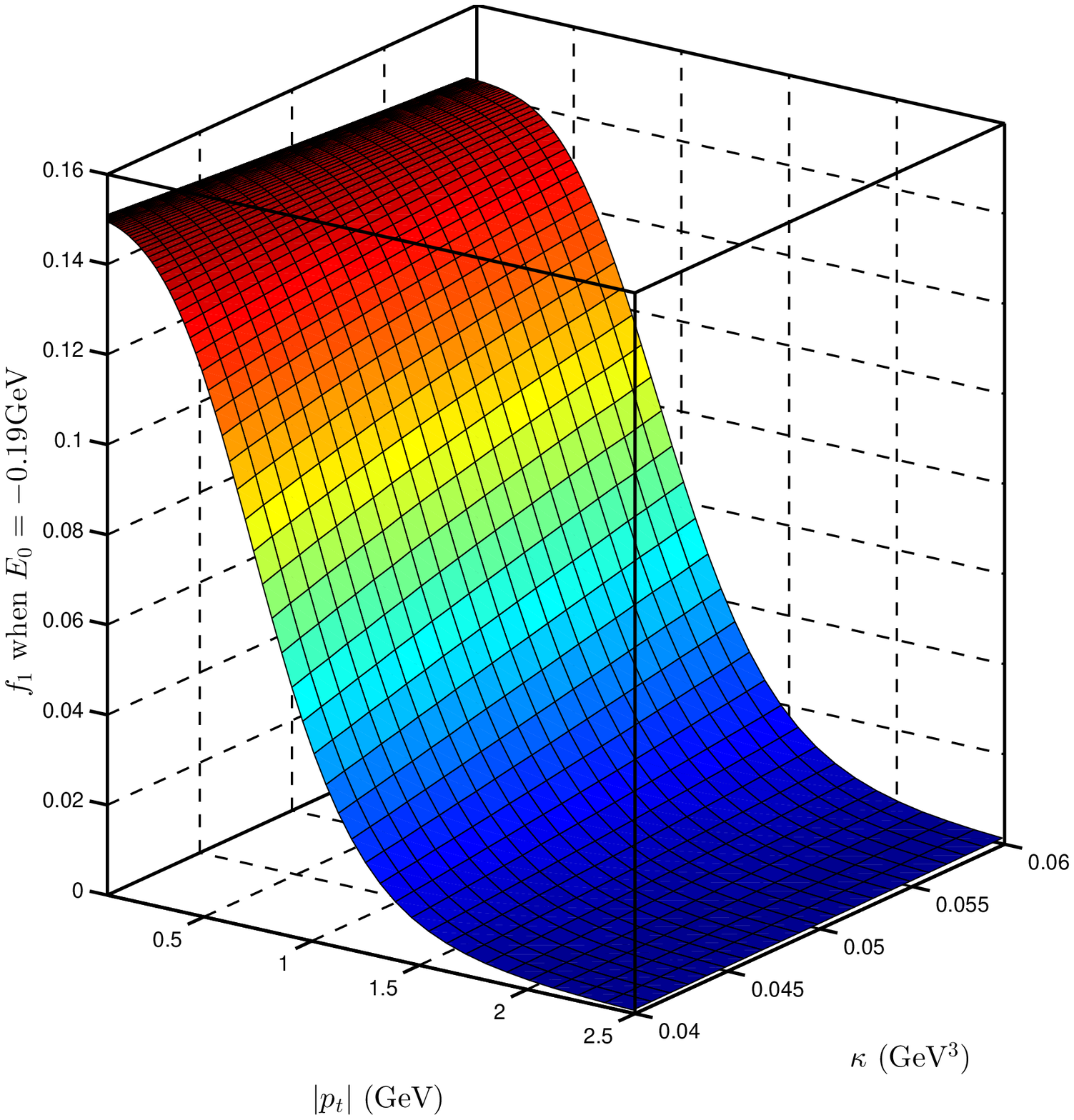}
\end{minipage}
\begin{minipage}[t]{0.45\linewidth}
 \includegraphics[width=7.0cm]{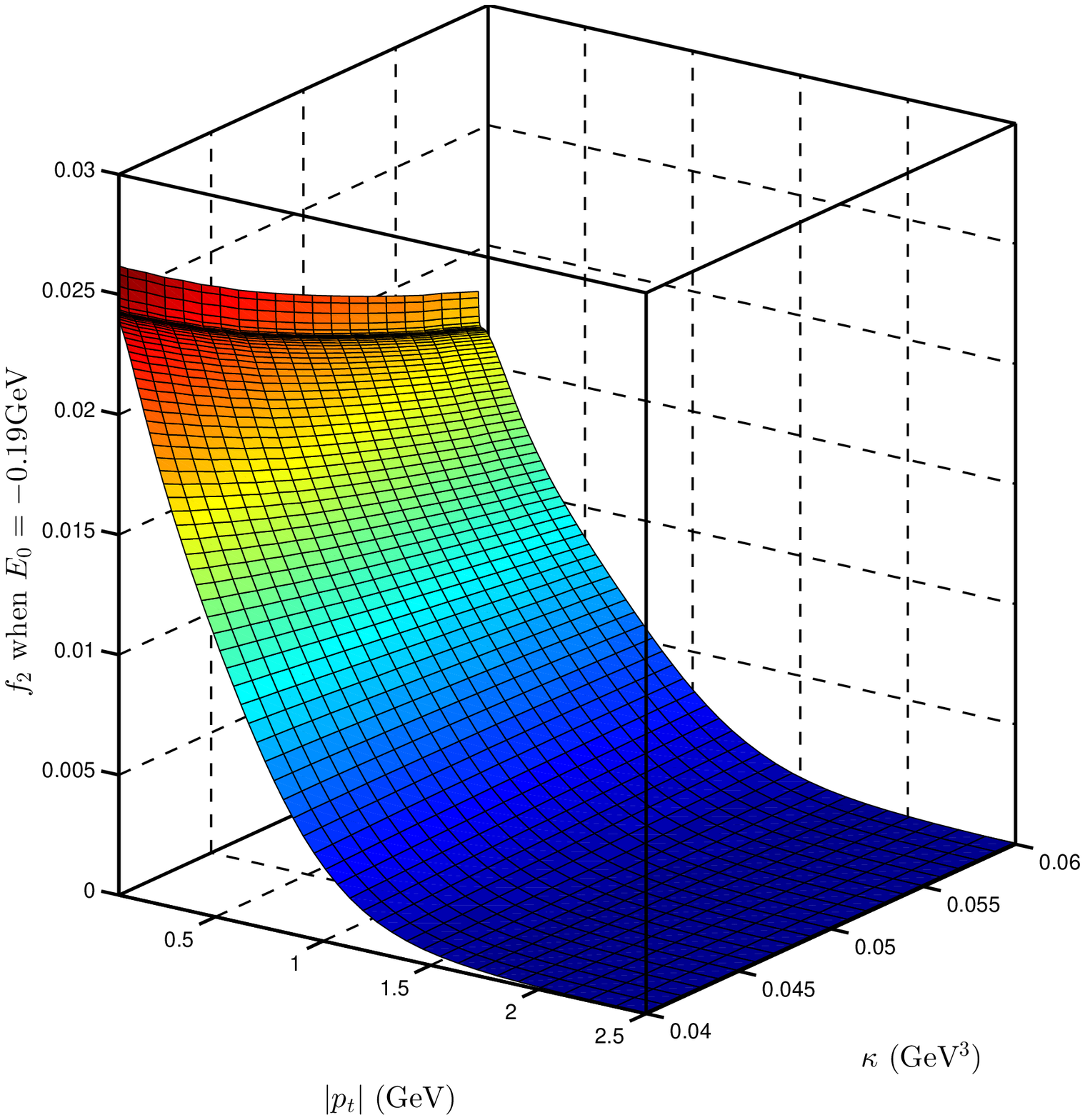}\label{GM75}
\end{minipage}
\caption{(color online) The BS wave functions for $\Lambda$ when $E_0=-0.19$ GeV.}\label{Fig:s19}
\end{center}
\end{figure}

\begin{figure}[!htb]
\begin{center}
\begin{minipage}[t]{0.45\linewidth}
 \includegraphics[width=7.0cm]{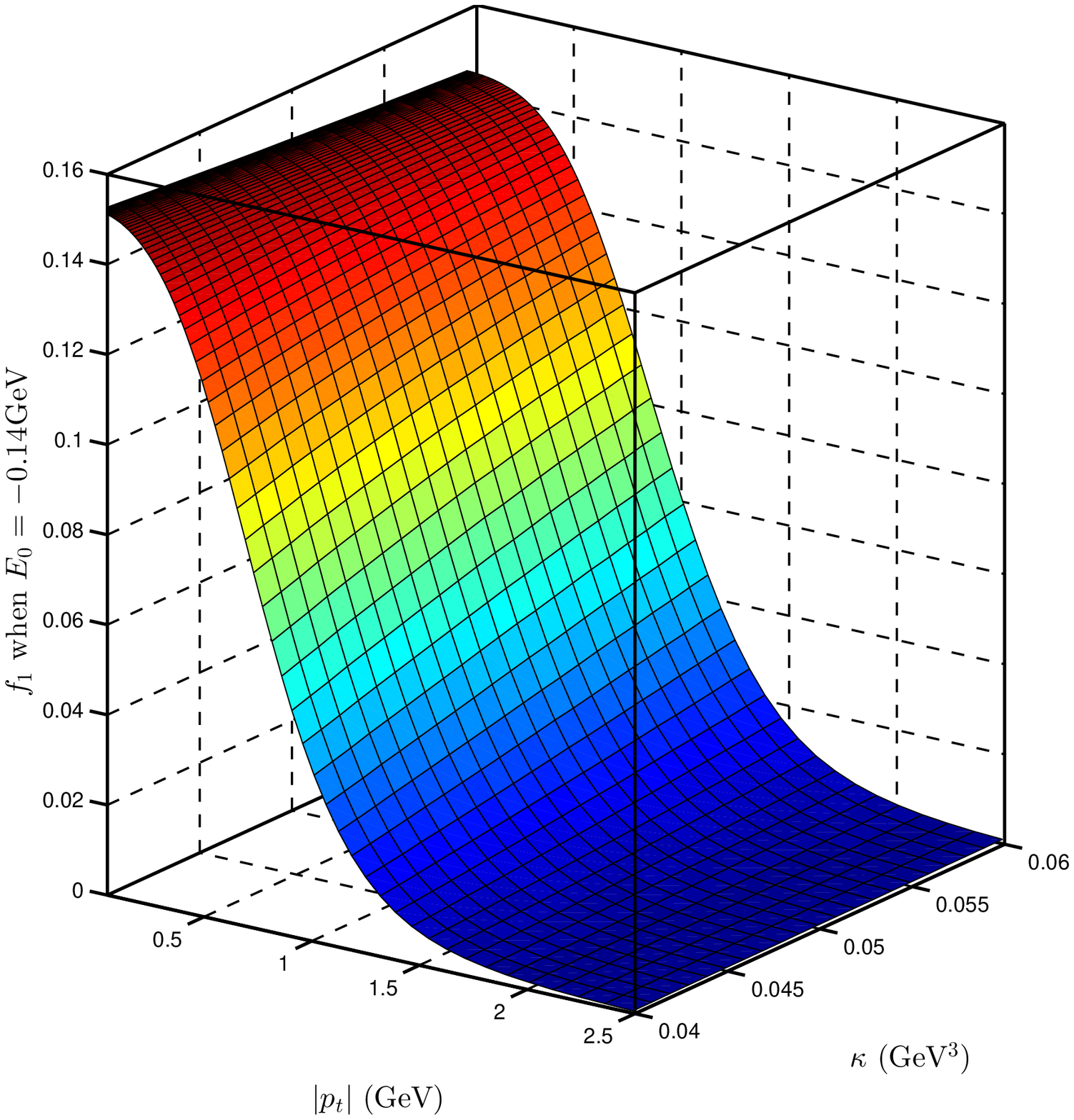}
\end{minipage}
\begin{minipage}[t]{0.45\linewidth}
 \includegraphics[width=7.0cm]{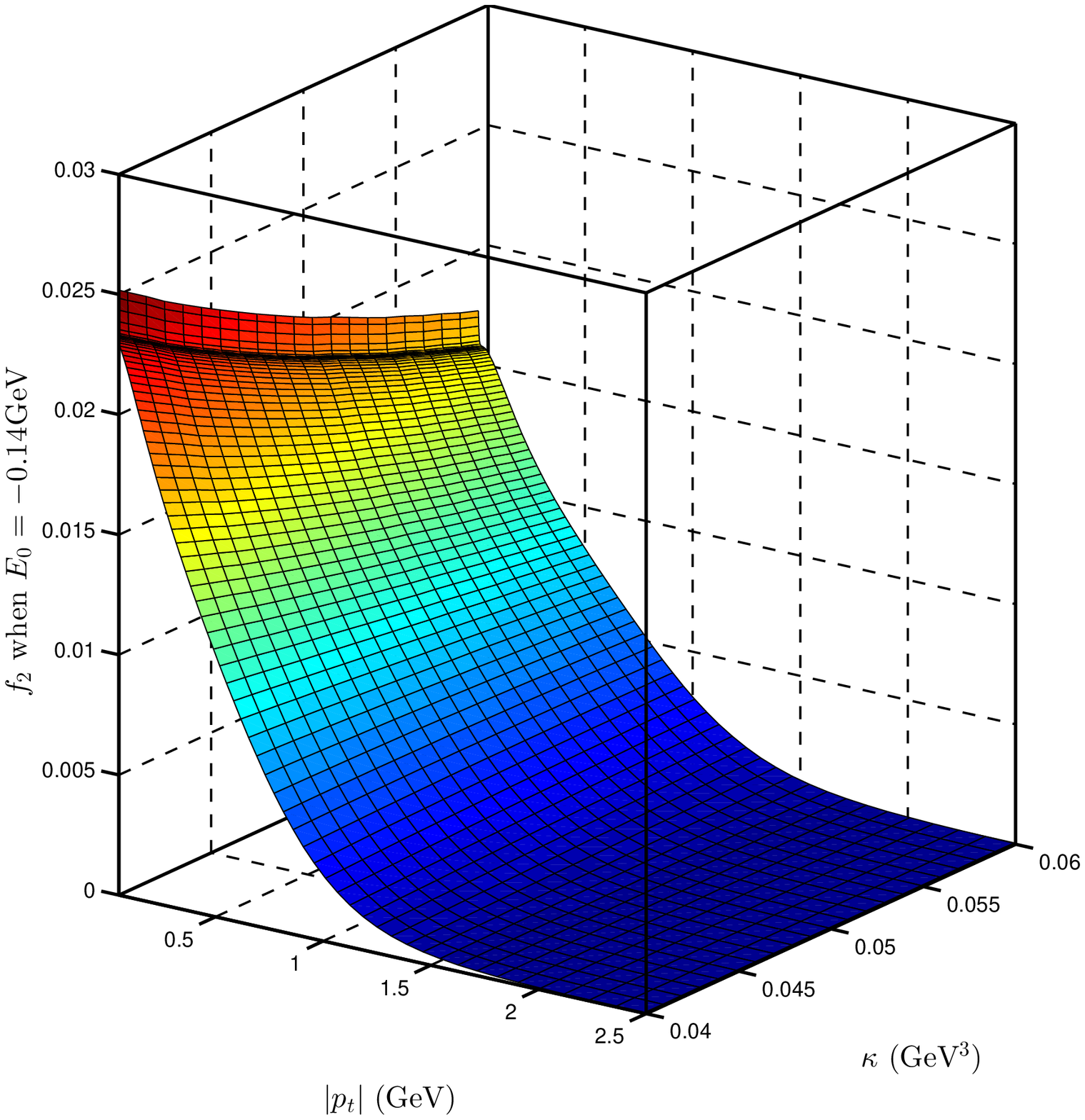}\label{GM75}
\end{minipage}
\caption{(color online) The BS wave functions for $\Lambda$ when $E_0=-0.14$ GeV.}\label{Fig:s14}
\end{center}
\end{figure}

\begin{figure}[!htb]
\begin{center}
\begin{minipage}[t]{0.45\linewidth}
 \includegraphics[width=7.0cm]{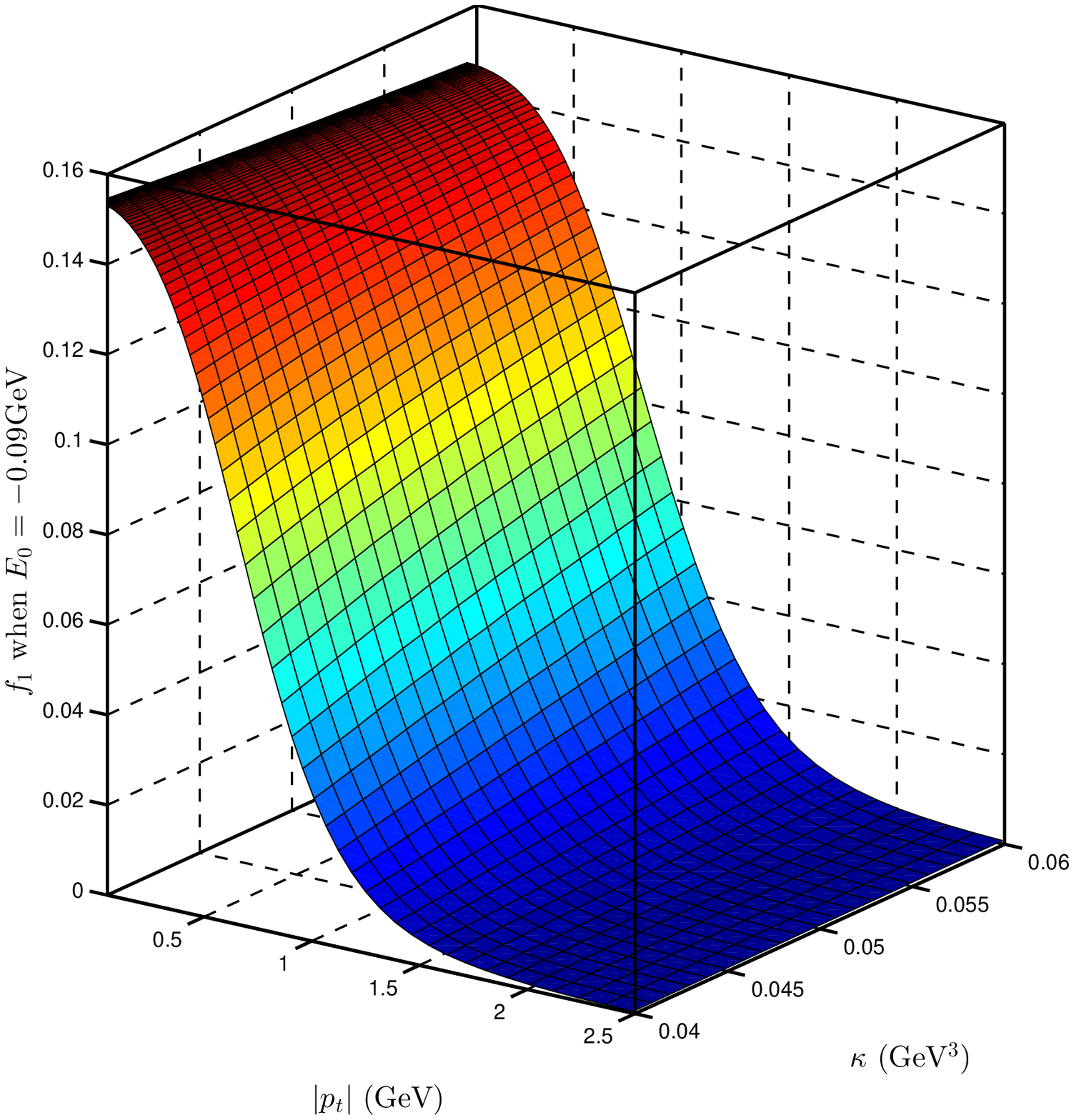}
\end{minipage}
\begin{minipage}[t]{0.45\linewidth}
 \includegraphics[width=7.0cm]{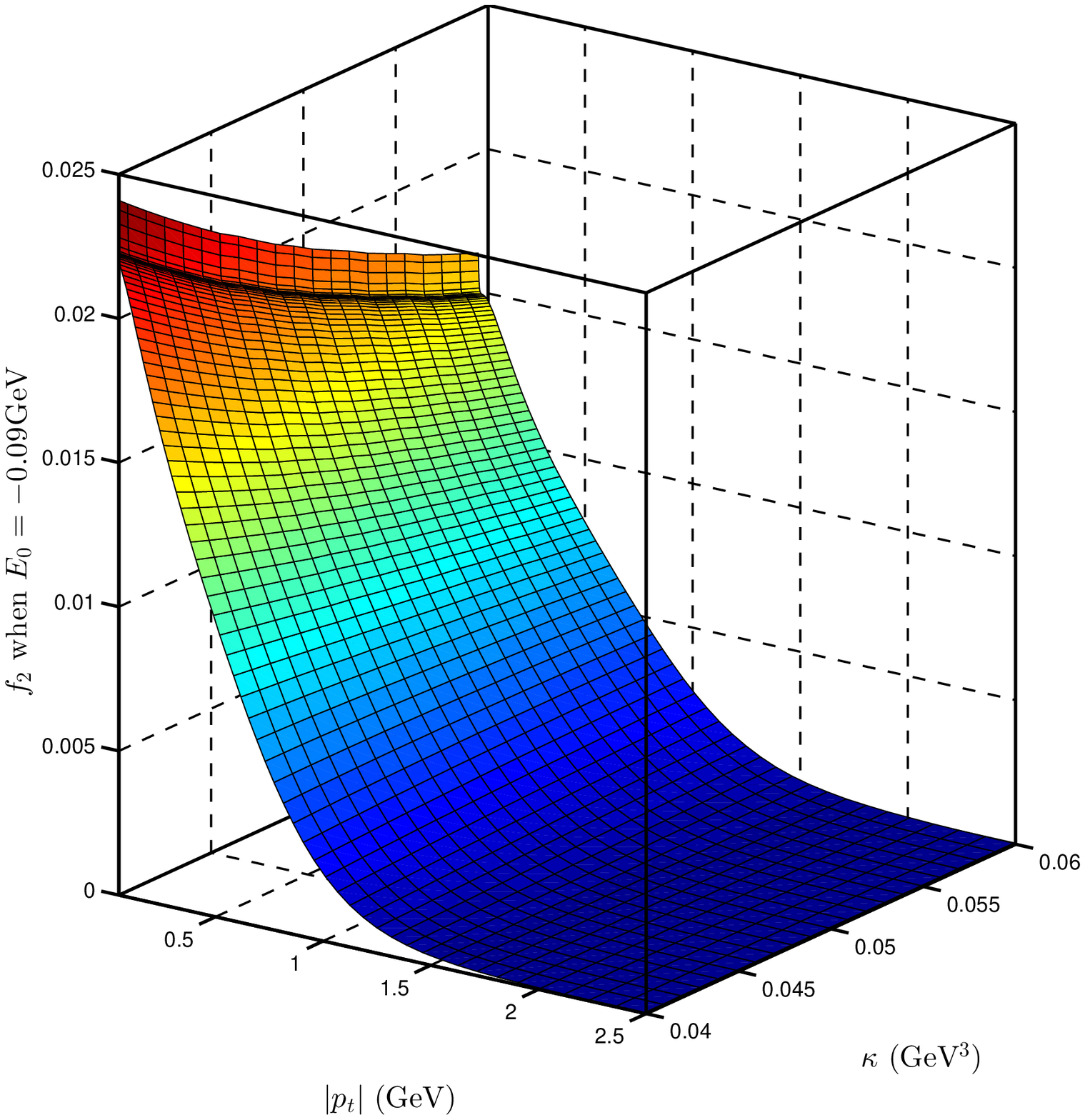}\label{GM75}
\end{minipage}
\caption{(color online) The BS wave functions for $\Lambda$ when $E_0=-0.09$ GeV.}\label{Fig:s9}
\end{center}
\end{figure}

\begin{figure}[!htb]
\begin{center}
\begin{minipage}[t]{0.45\linewidth}
 \includegraphics[width=7.0cm]{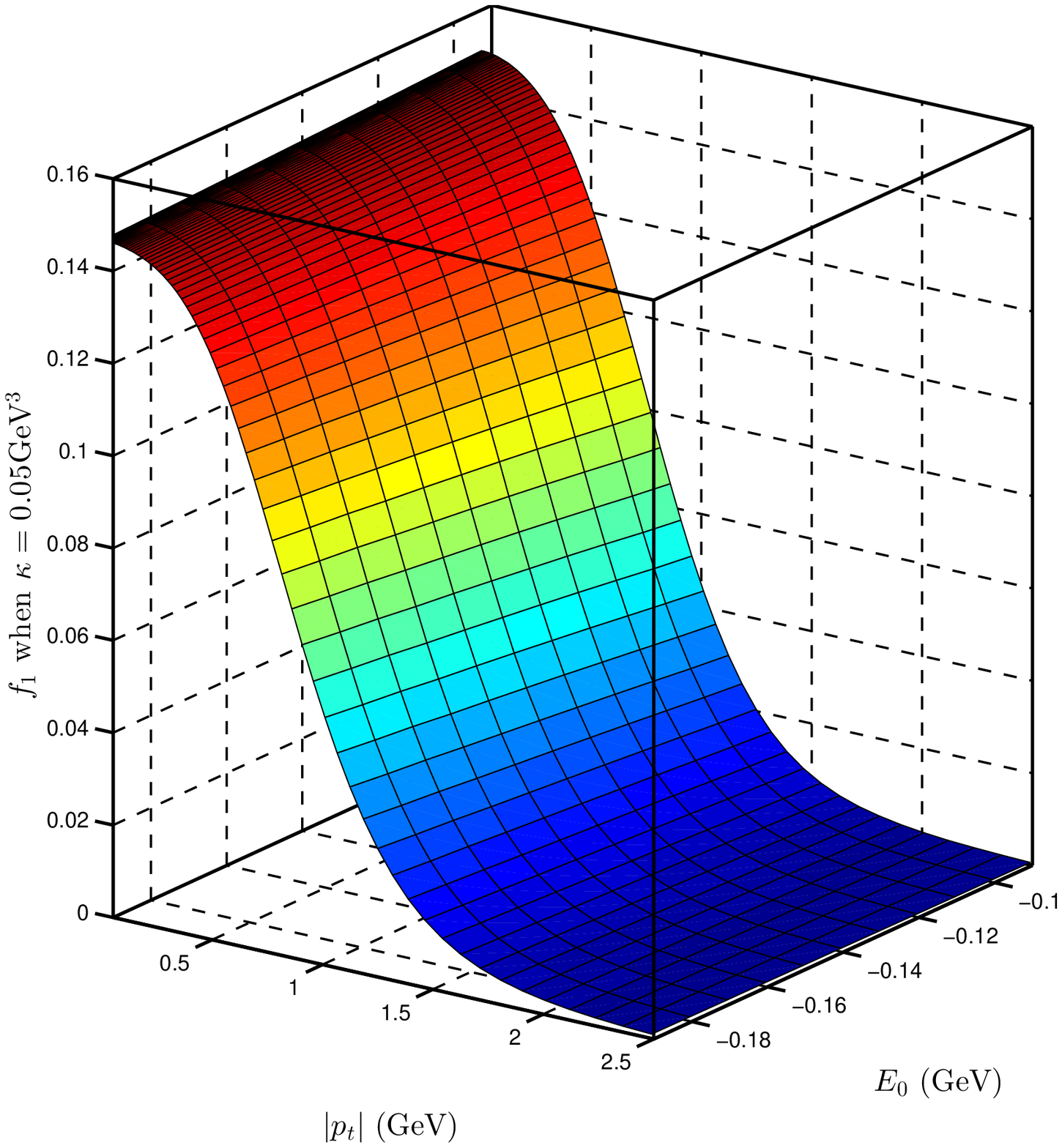}
\end{minipage}
\begin{minipage}[t]{0.45\linewidth}
 \includegraphics[width=7.0cm]{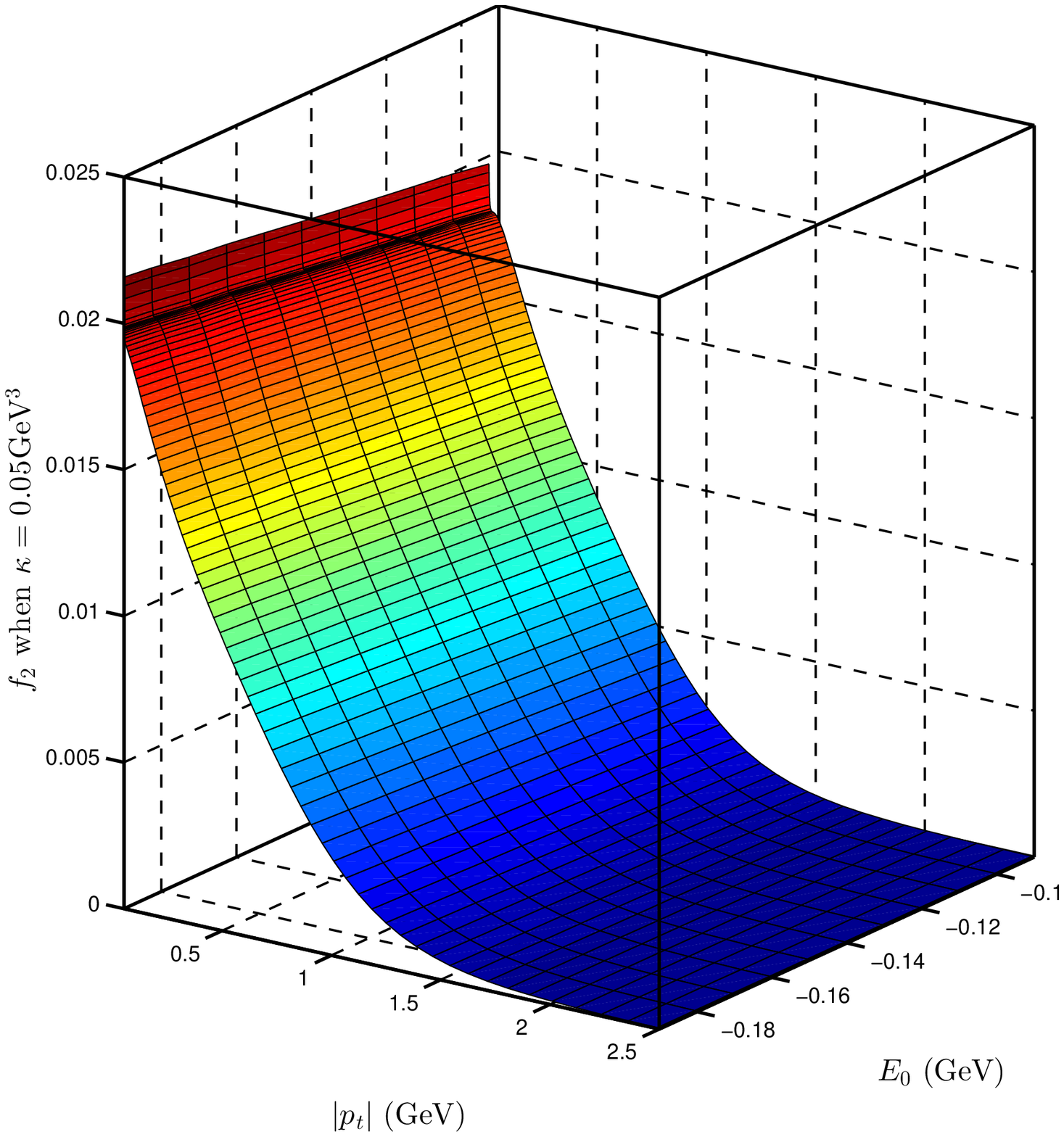}\label{GM75}
\end{minipage}
\caption{(color online) The BS wave functions for $\Lambda$ when $\kappa=-0.05$ GeV$^3$..}\label{Fig:s5}
\end{center}
\end{figure}

\begin{figure}[!htb]
\begin{center}
\begin{minipage}[t]{0.45\linewidth}
 \includegraphics[width=7.0cm]{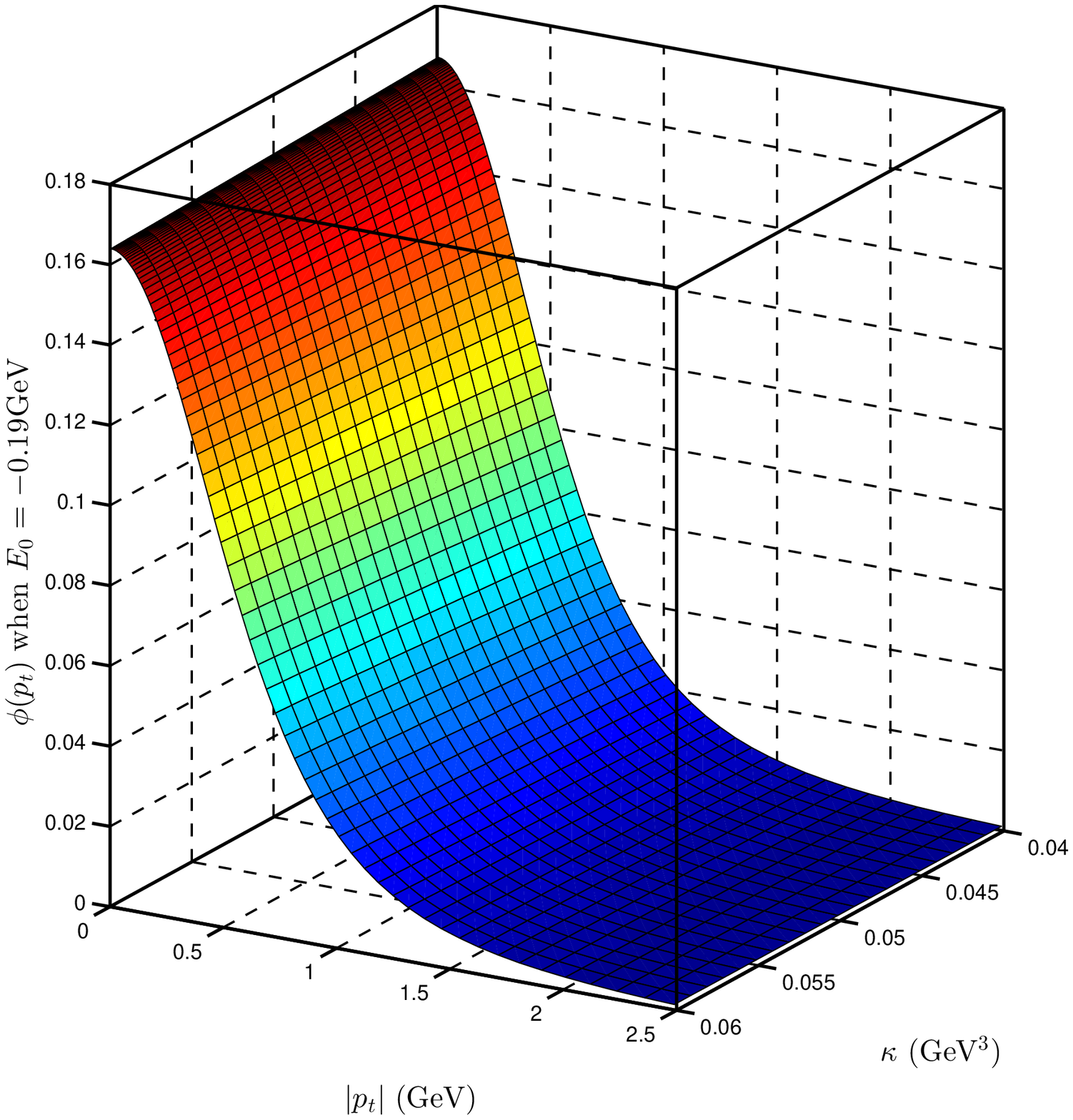}\label{Mf19kappa}
\end{minipage}
\begin{minipage}[t]{0.45\linewidth}
 \includegraphics[width=7.0cm]{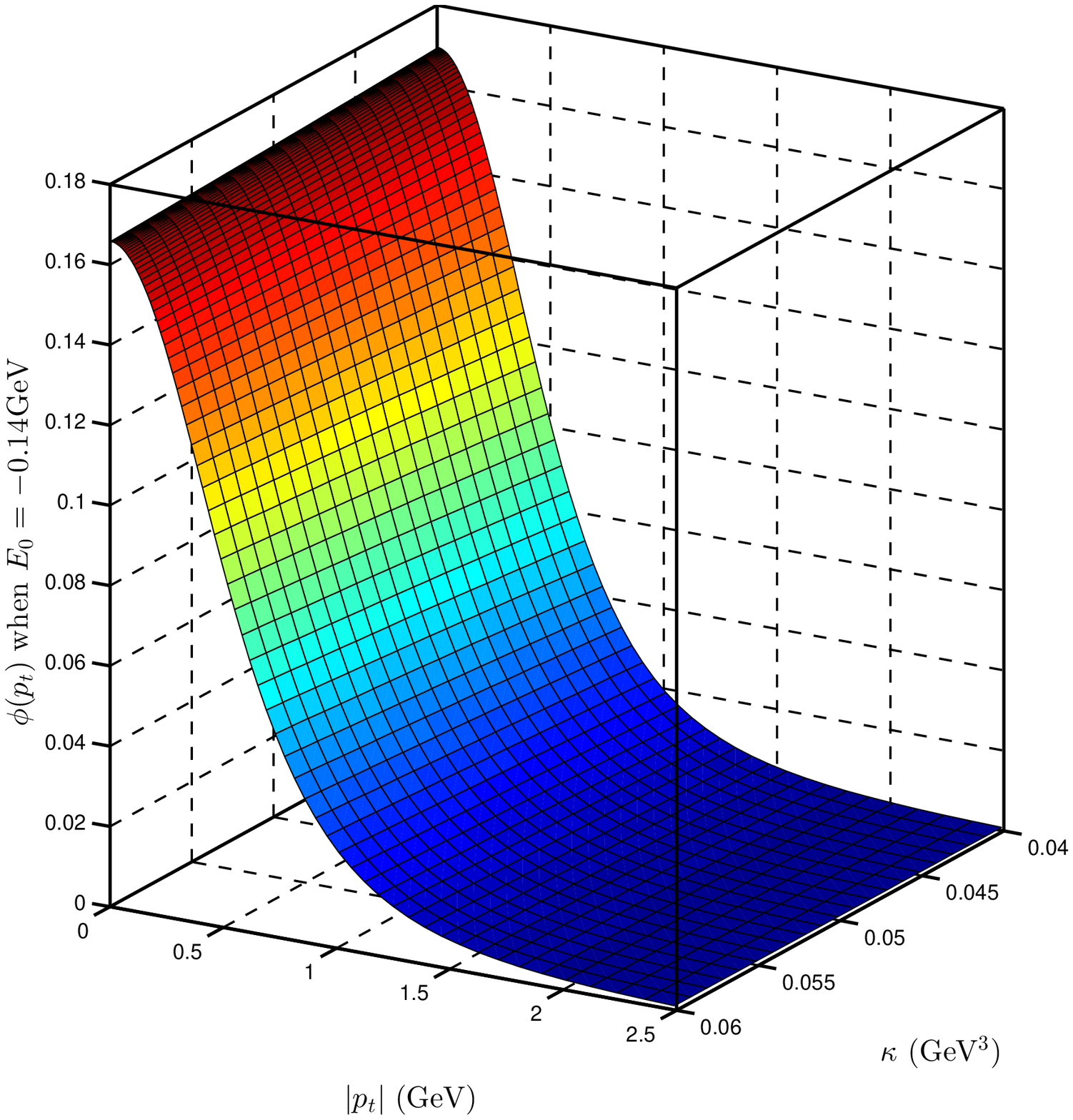}\label{Mf14kappa}
\end{minipage}
\caption{(color online) The BS wave function for $\Lambda_b$ when $E_0=-0.19$ GeV, $E_0=-0.14$ GeV.}\label{Fig:b19}
\end{center}
\end{figure}

\begin{figure}[!htb]
\begin{center}
\begin{minipage}[t]{0.45\linewidth}
 \includegraphics[width=7.0cm]{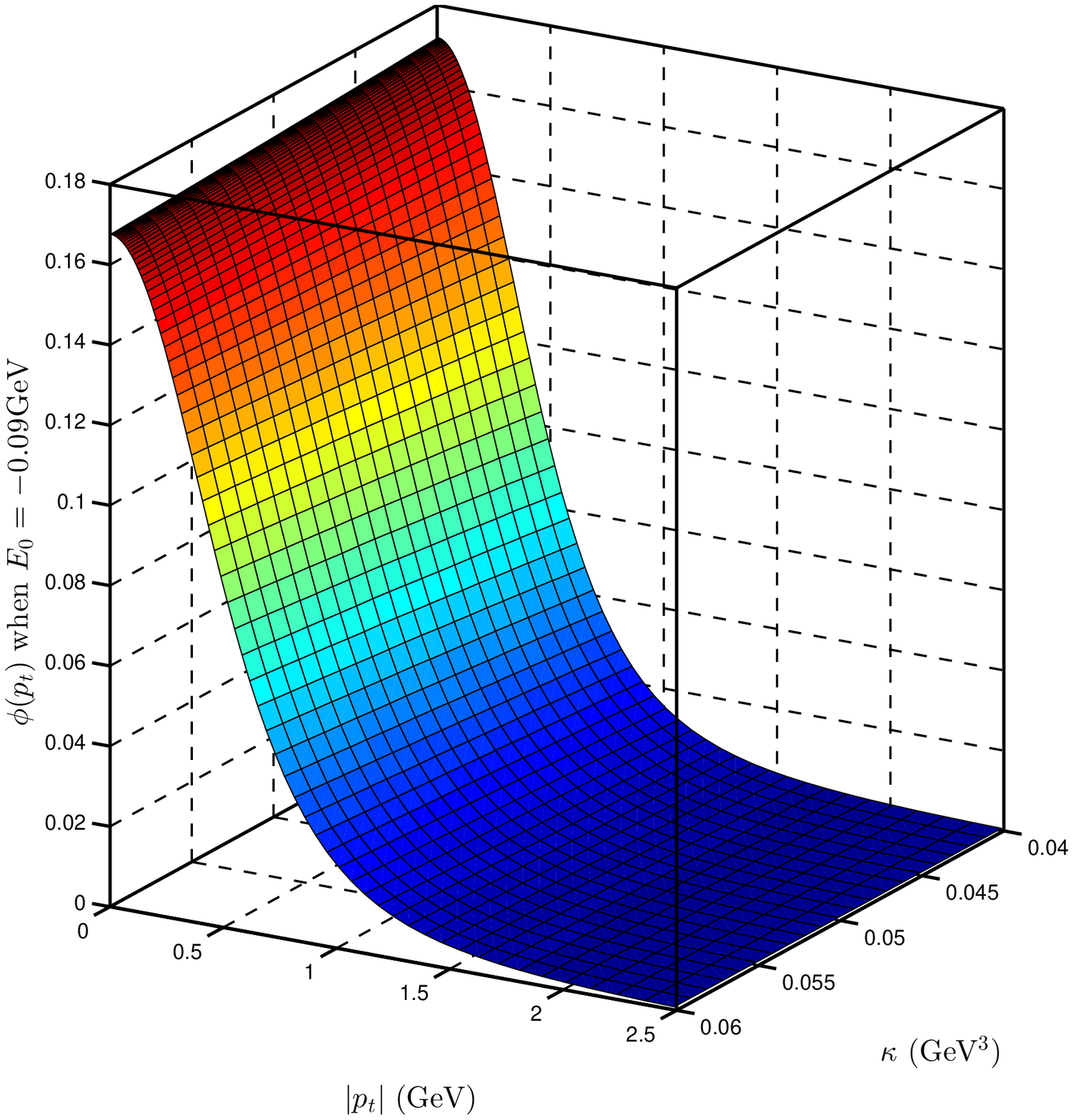}\label{Mf09kappa}
\end{minipage}
\begin{minipage}[t]{0.45\linewidth}
 \includegraphics[width=7.0cm]{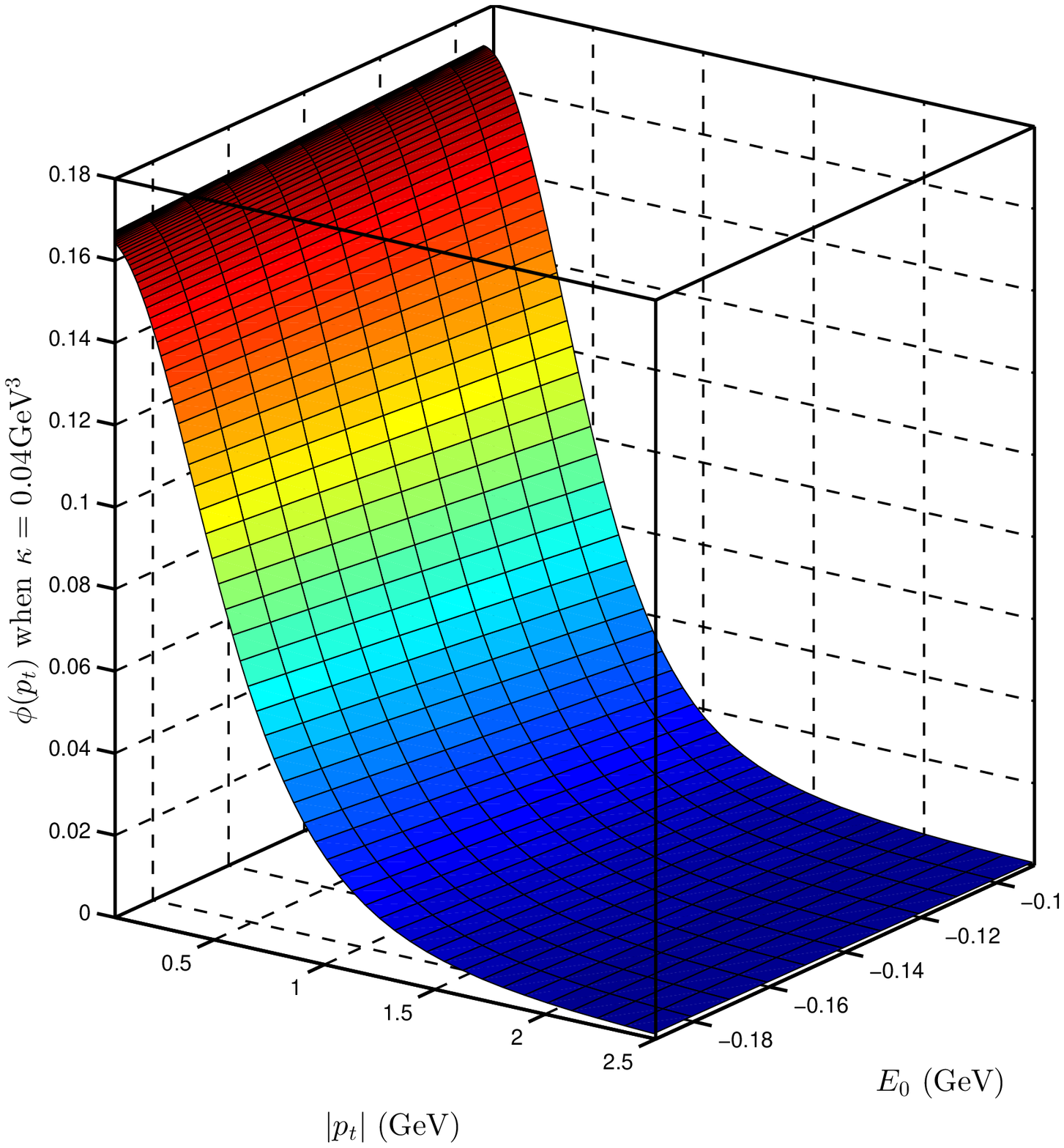}\label{Mf003e0}
\end{minipage}
\caption{(color online) The BS wave function for $\Lambda_b$ when $E_0=-0.09$ GeV, and $\kappa=-0.05$ GeV$^3$.}\label{Fig:b5}
\end{center}
\end{figure}

\begin{figure}[!htb]
  \centering
  \includegraphics[width=8.0cm]{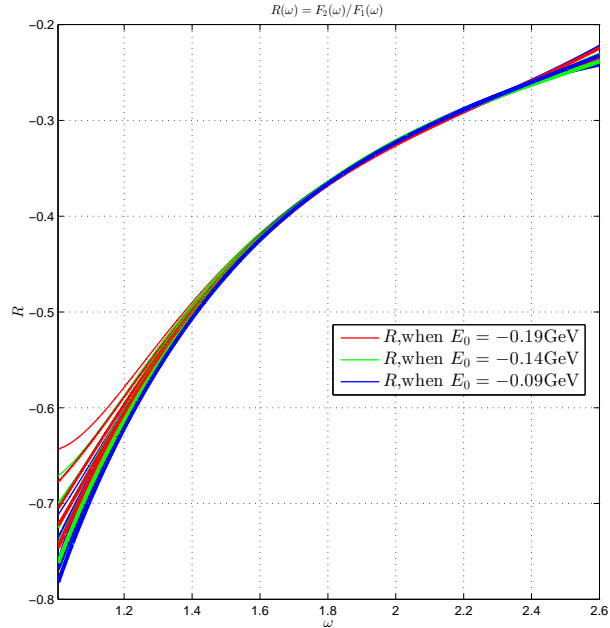}
  \caption{ (color online) The Values of $R(\omega)$ with different binding energy $E_0 $ and $\kappa$ (the values of $R$ decreases with the increases value of $\kappa$, and with increases the values of $\kappa$ the line gets thicker($\kappa$ from $0.040$ to $0.060$) for the same color line)}\label{FFR}
\end{figure}

In Figs. \ref{Fig:s19}- \ref{Fig:s5}, and Figs. \ref{Fig:b19}-\ref{Fig:b5}, we give the BS wave functions of $\Lambda$ and $\Lambda_b$ for different parameters.
From the figures in Figs. \ref{Fig:s19}- \ref{Fig:s5}, we find that the BS wave functions of $\Lambda$ are very similar for different parameters, the value of $f_1(\omega)$ changes from $0$ to about $0.15$, while the value of $f_2(\omega)$ changes from $0$ to about $0.022$.
However, $f_2(\omega)$ depends on $\kappa$ more heavily than on $E_0$.
From the figures in Figs. \ref{Fig:b19}-\ref{Fig:b5}, we find that the BS wave functions of $\Lambda_b$ are very similar for different parameters.
In Figs. \ref{FFR}, we give the values of $R(\omega)$ for different parameters.
From this figure, we find that the values of $R(\omega_{max})$ ($=R(q^2=0)$) are all about $-0.23$ for different parameters, this value agrees with the experimental result very well \cite{PRL75-624}.
The value of $R$ varies from $-0.8$ to $-0.23$ for different $E_0$ and $\kappa$ when $\omega = 1 \sim 2.6$ (corresponding to $q^2$ from $ m^2_e $ to $(m_{\Lambda_b}-m_\Lambda)^2 $ ).
This range agrees with our result and that in Ref. \cite{PRD59-114022}.
Considering the experimental data for $R(\omega)$ in Ref. \cite{PRL94-191801} and the values of $R(\omega=1)$ decreases with the increase of values of $\kappa$ or $E_0$, we believe that the optimal range for our model parameters is $\kappa=0.050$ GeV$^3$ and $E_0$ from $-0.19$ to $-0.09$ GeV, because in this region $R(q^2_{max})=-0.8\sim -0.7$ and $R$ varying from $-0.8$ to $-0.23$ agree with our previous results.
On the other hand, we find that LQCD also gives the value $R(q^2_{max}) \approx -0.8$ \cite{PRD87-074502}.

\begin{table}[!htb]
\centering  
\begin{tabular}{c||c|c|c|c|c}  
\hline
\makecell[c]{  \\ $-E_0 $($\times 10^2$GeV) \\ $\kappa$($\times 10^3$GeV$^3$)} &\makecell[c]{ present work\\ $ 14$ \\ $ 50 \pm 5 $} &\makecell[c]{ present work\\ $ 14\pm5 $ \\ $ 50 $} & HQET\cite{PRD64-074001} & QCD sum rules \cite{PLB516-327} & Exp.\cite{PRD98-030001}  \\ \hline \hline
$Br( \Lambda_b\rightarrow \Lambda e^+ e^- ) \times 10^{6}$&0.464-1.144&0.611-0.867 &2.23-3.34 &4.6$\pm$1.6 &- \\         
$Br( \Lambda_b\rightarrow \Lambda \mu^+ \mu^- )\times 10^{6}$&0.602-1.482&0.856-1.039 &2.08-3.19 &4.0$\pm$1.2 &1.08$\pm$0.28 \\
$Br( \Lambda_b\rightarrow \Lambda \tau^+ \tau^- )\times 10^{6}$&0.177-0.437& 0.233-0.331&0.179-0.276 &0.8$\pm$0.3 &- \\
 \hline \hline
\end{tabular}
\caption{The values of the branching ratios of $\Lambda_b\rightarrow \Lambda l^+ l^-$ and compare with other model.}\label{TB3}
\end{table}

\begin{figure}[!htb]
  \centering
  \includegraphics[width=8.0cm]{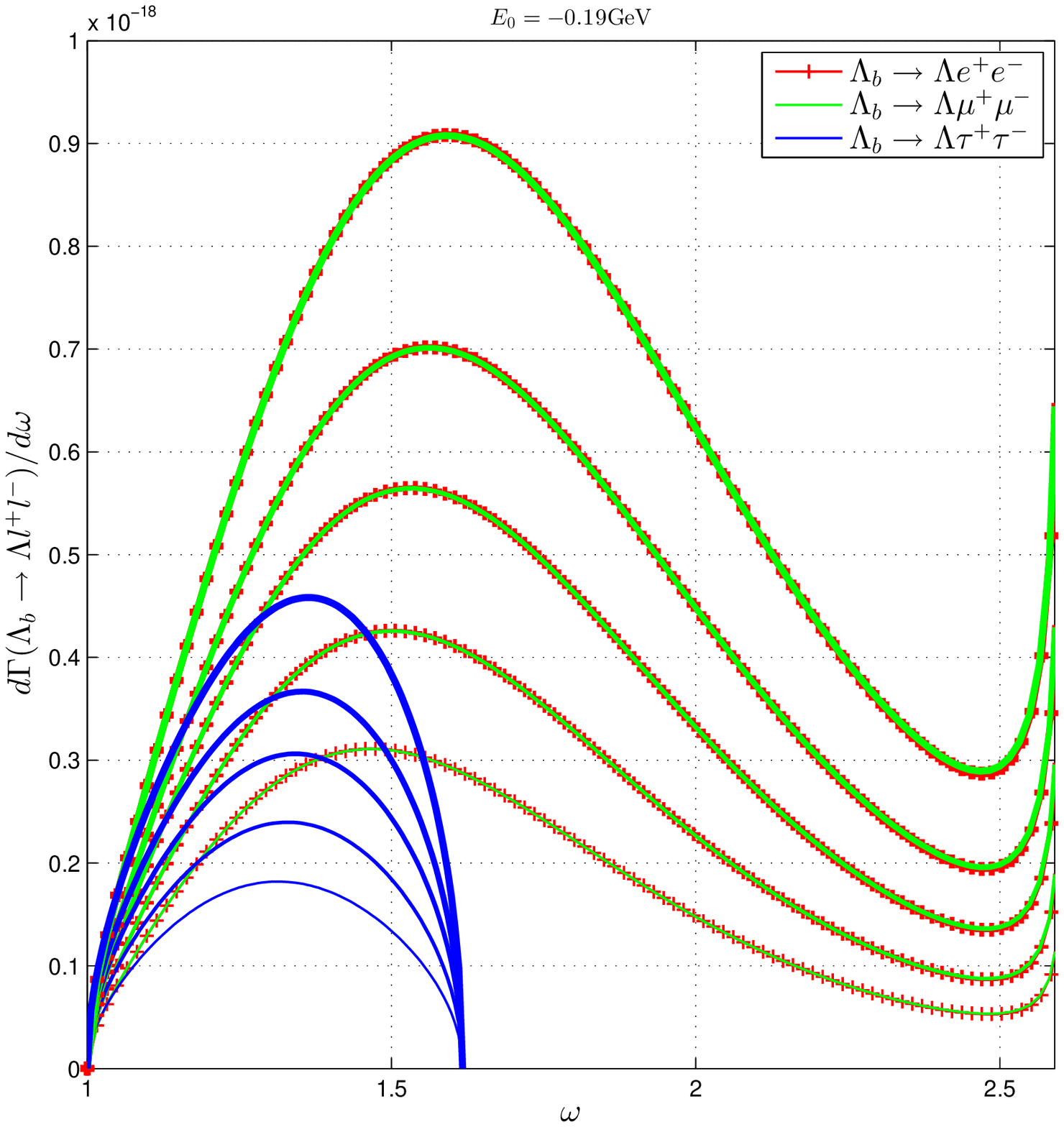}
  \caption{(color online) The differential decay width of $\Lambda_b \rightarrow \Lambda l^+ l^-$ when binding energy $E_0=-0.19$ GeV (the values of decay width increases with the increases value of $\kappa$ from $0.040$ to $0.060$ GeV$^3$) for the same color line ).}\label{DW:19}
\end{figure}

\begin{figure}[!htb]
  \centering
  \includegraphics[width=8.0cm]{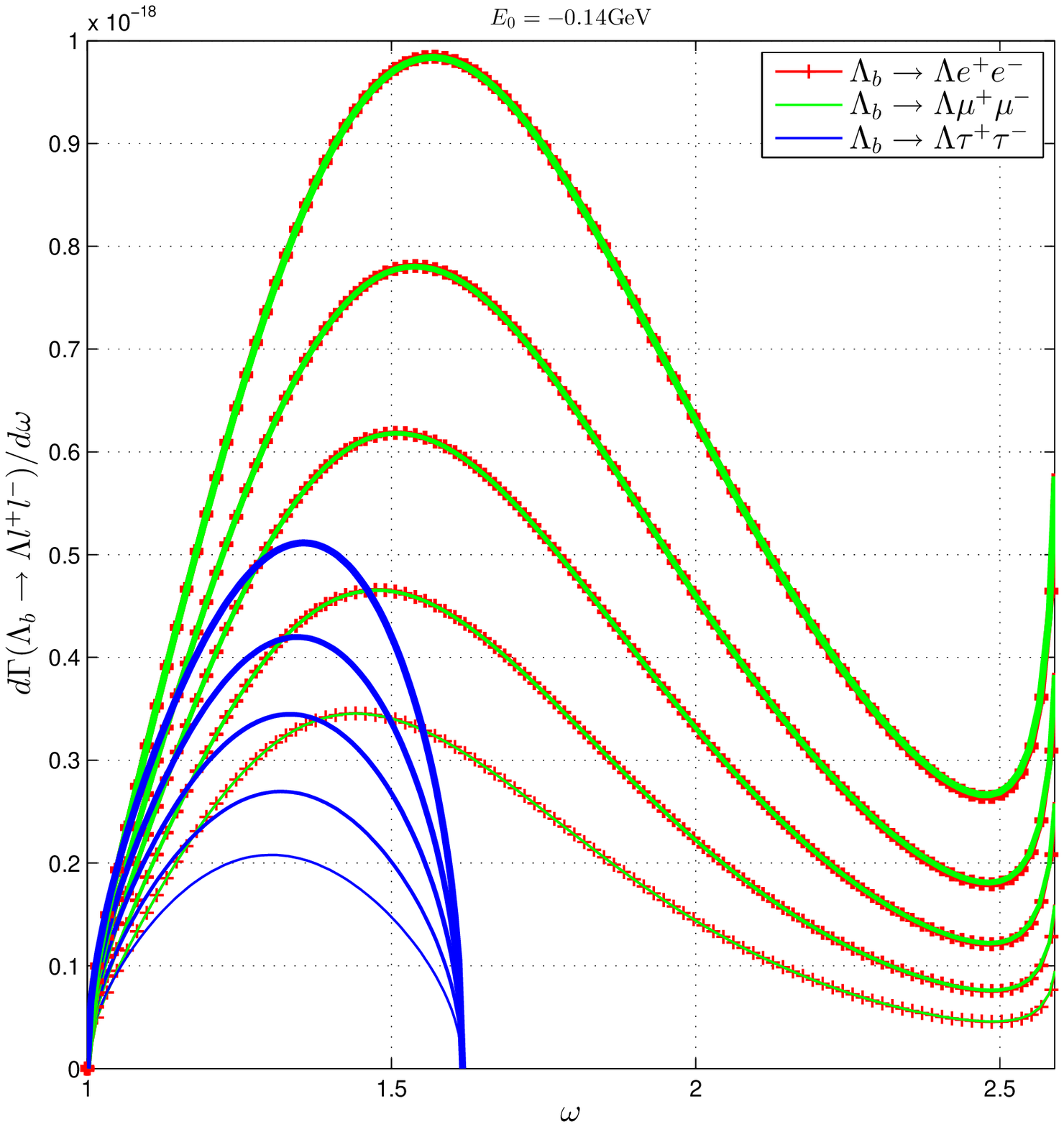}
  \caption{(color online) The differential decay width of $\Lambda_b \rightarrow \Lambda l^+ l^-$ when the binding energy $E_0=-0.14$ GeV (the decay width increases with the increase $\kappa$ from $0.040$ to $0.060$ GeV$^3$) for the same color line).}\label{DW:14}
\end{figure}

\begin{figure}[!htb]
  \centering
  \includegraphics[width=8.0cm]{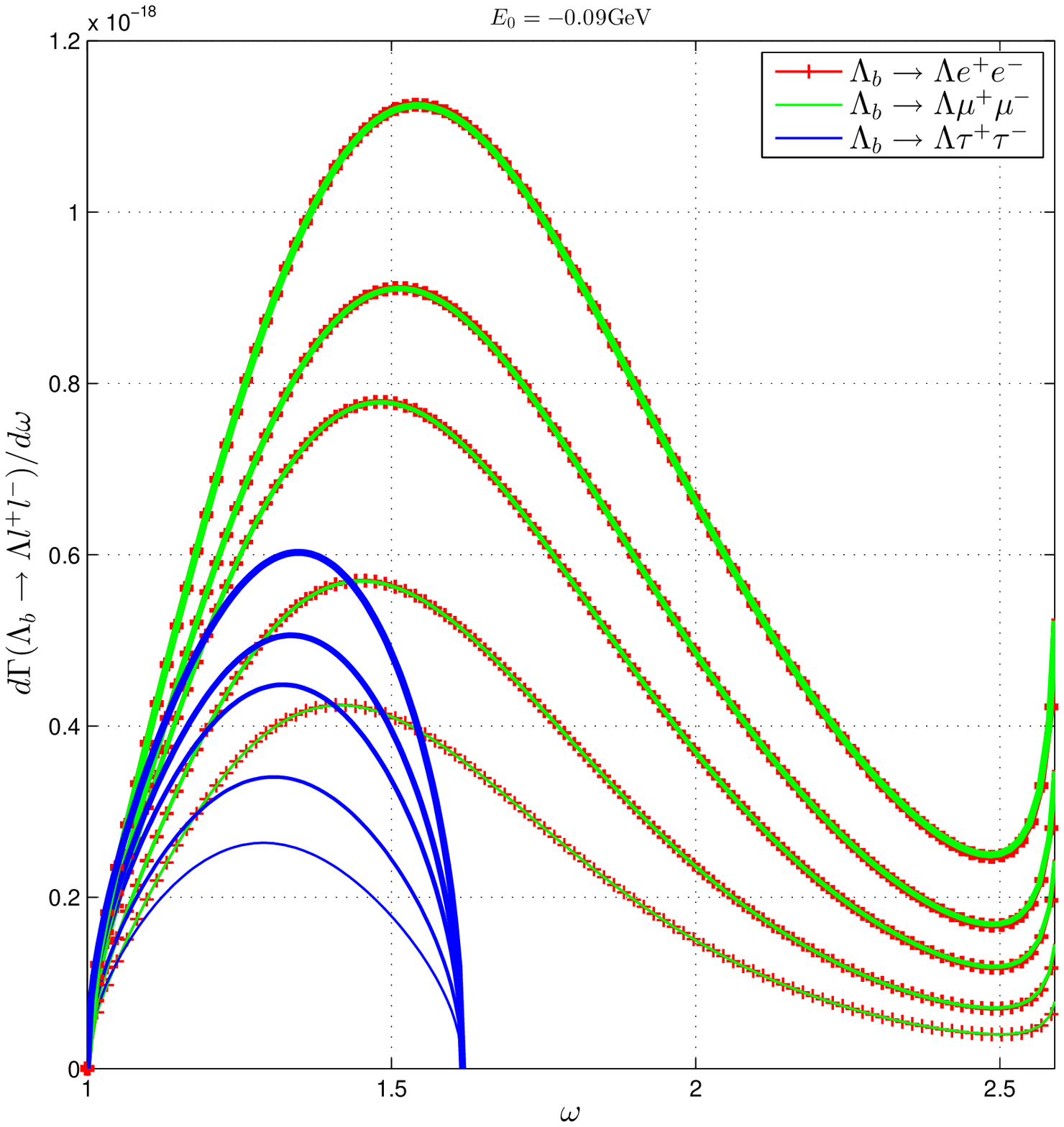}
  \caption{(color online) The differential decay width of $\Lambda_b \rightarrow \Lambda l^+ l^-$ when the binding energy $E_0=-0.09$ GeV (the decay width increases with the increase of $\kappa$ from $0.040$ to $0.060$ GeV$^3$) for the same color line)}\label{DW:09}
\end{figure}

In Figs. \ref{DW:19}-\ref{DW:09}, we give the $\omega$-dependent differential decay width of $\Lambda_b \rightarrow \Lambda l^- l^+(l= e, \mu, \tau)$ for different parameters.
In our optimal range of parameters and in the range $\kappa=0.050 \pm 0.005$ GeV$^3$, and $E_0=-0.14 \pm 0.5$GeV, we obtain the branching ratios, respectively, which are listed in Table \ref{TB3}.
From this table, we can see that our results are different from those of HQET and QCD sum rules, but our results are consistent with the most recent experimental data.
When $\kappa=0.045 \sim 0.055$ GeV$^3$ and $E_0=-0.19 \sim -0.14$ GeV, we find $Br(\Lambda_b \rightarrow \Lambda \mu^+ \mu^-)\times 10^6= 0.602 \sim 1.48$, and in our optimal parameter range this value is $0.856\sim 1.039$.
The values of $Br(\Lambda_b \rightarrow \Lambda e^+(\tau^+) e^-(\tau^-))\times 10^6$ in the above two ranges are  $0.464\sim1.144~(0.611\sim0.867)$ and $0.177\sim0.437~(0.233\sim0.331)$, respectively.
When the parameters $\kappa$ and $E_0$ vary in their regions, we find that the differential branching ratio of $\Lambda_b \rightarrow \Lambda \mu^+ \mu^-$ does not have a pole at about $\omega=1.2$.
In Refs. \cite{PLB725-23,JHEP06-115} when $\omega$ is in the range $1 \sim 1.4$ (corresponding to $q^2$ in the range $15 \sim 20 $GeV$^2$), the experimental data have a pole.
Considering this different, there could be new physics in this region.

\section{summary and discussion}

Theoretical studies of the decay $\Lambda_b \rightarrow \Lambda l^+ l^-$ require knowledge of the matrix element $\langle \Lambda| \bar{s} \Gamma b| \Lambda_b\rangle$.
At the leading order in the heavy quark effective theory, this matrix element is given by two FFs.
In the past few decades, in most of works the FFs were studied based on QCD sum rules \cite{PRD59-114022}, and by fitting the experimental data \cite{PRL75-624}.
With the progresses of experiments, the data about $\Lambda_b$ rare decay has been updated.
In the present work, we have performed the first BS equation calculation of these FFs.
In our work, $\Lambda_Q~(Q=b,s)$ is regarded as a bound state of a $Q$-quark and a scalar diquark.
In this picture, we established the BS equations for $\Lambda_Q$, and derived the FFs for $\Lambda_b \rightarrow \Lambda $ in the BS equation approach.
After solving the BS equations of $\Lambda$ and $\Lambda_Q$.
We calculated  the value of $R$, and decay branching ratio for $\Lambda_b \rightarrow \Lambda l^+ l^-$ also compared our results with other theoretical works and the experimental data.
We found that the shapes of the differential decay branching ratio for $\Lambda_b \rightarrow \Lambda \mu^+ \mu^-$ in our model is similar to the experimental data in most part of the region and in our work the shapes of the decay differential branching ratio of $\Lambda_b \rightarrow \Lambda l^+ l^-(l= e, \mu, \tau)$ agree with those of LQCD \cite{PRD87-074502, PRD93-074501}.
The experimental data for the differential decay width of $\Lambda_b \rightarrow \Lambda \mu^+ \mu^-$ have a pole when $\omega \approx 1.2$, but in most of theoretical works such a pole does not appear.
Therefore, in this region there could be new physics.
The experimental data need to be improved for higher accuracy in remeasure this region.
Our result for $\Lambda_b \rightarrow \Lambda \mu^+ \mu^-$ is very close to the experimental data and we also give the predictions for the decays $\Lambda_b \rightarrow \Lambda l^+ l^-(l=e, \tau)$, which need to be tested in future experimental measurements.
We find that for different values parameters the FFs ratio $R(\omega)$ changes from $-0.80$ to $- 0.23$ in our approach.
This result agrees with the experimental data and that in Ref. \cite{PRD59-114022}, and agrees with LQCD at $q^2_{max}$ \cite{PRD87-074502}.
In the heavy quark effective theory, the approximation $1/m_b \rightarrow \infty$ leads to an uncertainty of about $\Lambda_{QCD}/m_b$.
Considering the uncertainties from the parameters $E_0$ and $ \kappa$ the maximum uncertainty is about $22\%$ in our optimal data region.

In the future, our model can also be used to study the forward-backward asymmetries, T violation and angular distributions in the decays induced by $ b \rightarrow s l^+ l^- $ to further check our FFs.

\acknowledgments
This work was supported by National Natural Science Foundation of China under contract numbers 11775024, 11575023, 11847052, 11981240361 and 11905117.


\end{document}